\documentclass[10pt,a4paper]{article}

\usepackage{natbib}
\usepackage{geometry}
\usepackage{amsmath}
\usepackage{amsthm}
\usepackage{amssymb}
\usepackage{verbatim}
\usepackage{booktabs}
\usepackage{tabularx}
\usepackage{setspace}
\usepackage[singlelinecheck=false]{caption}

\begin{document}

\title{Evaluating Callable and Putable Bonds: An Eigenfunction Expansion Approach\footnote{This research was supported by the National Science Foundation grants DMS-0802720 and DMS-1109506.}}

\author{Dongjae Lim\footnote{Department of Industrial Engineering and Management Sciences, McCormick School of Engineering and Applied Sciences, Northwestern University, 2145 Sheridan Road, Evanston, IL 60208, Email addresses: \texttt{ dongjae@u.northwestern.edu} (Dongjae Lim), \texttt{ lingfeili2012@u.northwestern.edu} (Lingfei Li), \texttt{ linetsky@iems.northwestern.edu} (Vadim Linetsky) } \and Lingfei Li\footnotemark[\value{footnote}] \and Vadim Linetsky\footnotemark[\value{footnote}]}

\date{June 1, 2012}

\maketitle
\begin{abstract}
We propose an efficient method to evaluate callable and putable bonds under a wide class of interest rate models, including the popular short rate diffusion models, as well as their time changed versions with jumps. The method is based on the eigenfunction expansion of the pricing operator. Given the set of call and put dates, the callable and putable bond pricing function is the value function of a stochastic game with stopping times. Under some technical conditions, it is shown to have an eigenfunction expansion in eigenfunctions of the pricing operator with the expansion coefficients determined through a backward recursion. For popular short rate diffusion models, such as CIR, Vasicek, 3/2, the method is orders of magnitude faster than the alternative approaches in the literature. In contrast to the alternative approaches in the literature that have so far been limited to diffusions, the method is equally applicable to short rate jump-diffusion and pure jump models constructed from diffusion models by Bochner's subordination with a L\'{e}vy subordinator. 
\mbox{} \\

\mbox{} \\

\end{abstract}

\newtheorem{thm}{Theorem}[section]
\newtheorem{lem}[thm]{Lemma}
\newtheorem{propo}[thm]{Proposition}
\theoremstyle{remark}
\newtheorem{remark}{Remark}[section]
\renewenvironment{proof}[1][\proofname]{{\noindent \bfseries #1.  }}{\qed\\}

\section{Introduction}

A large fraction of all corporate and sovereign bond issues in the global financial markets have embedded options. The call option allows the bond issuer, such as a corporation or a government, to buy the bond back from the bond holder (call the bond) for pre-specified call prices at some pre-specified times prior to bond's maturity. This allows the bond issuer to refinance the bond if interest rates decline.  The put option allows the bond holder to sell (put) the bond back to the bond issuer for pre-specified put prices at some pre-specified times prior to maturity. This allows the bond holder to re-invest the proceeds into a bond with a higher coupon if interest rates rise. 
The bond with both a call and a put option can be analyzed as an instance of a stochastic game with stopping times (also known as Dynkin games as they have been introduce by \citet{dynkin} as a generalization of optimal stopping problems) driven by the  
underlying stochastic interest rate model.
The bond issuer and the bond holder are opposing players whose opposing optimal strategies are to minimize and to maximize the bond value, respectively. When the call and put decisions can be made at discrete times (typically an advance notice has to be given to the other party ahead of a coupon payment date, when the option is exercised), this sets up a stochastic game with stopping times in discrete time. The value function and the optimal call and put policies can then be determined by solving Bellman's dynamic programming backward induction, starting from maturity and rolling back recursively through the decision times. 
Developing solution methods for this problem is of significant practical importance.

The pricing of bonds with embedded options has attracted considerable interest in the literature over the years.
It goes back to \cite{brennan} who used a finite-difference approach in time-homogeneous diffusion models.
Later \cite{buttler} showed that when finite-difference methods were used to price callable bonds under the Vasicek model, the presence of slowly decaying oscillations in the solution after each coupon/call date resulted in poor numerical accuracy.  
This led \cite{bw} (BW) to develop an alternative method 
for pricing callable bonds under the Vasicek and CIR models utilizing the explicit form of the Green's function in these models.
The method relies on the interpolation of the value function and on the numerical quadrature procedure for the integration involving the value function and the Green's function. More recently \cite{dfvl} (DFVL) showed that finite-difference methods for these problems can be stabilized via van Leer flux limiter, appropriate non-uniform time stepping schemes, and careful consideration of boundary conditions and presented numerical experiments demonstrating that properly formulated finite-difference methods, in fact, outperformed alternative approaches. Other recent works on applying numerical PDE methods to callable and putable bonds in diffusion interest rate models include   \cite{farto} and \cite{frutos}. 
In \cite{farto}, the convection dominated diffusion equation is solved numerically by combining the characteristics method with piecewise-linear Lagrange finite elements.  \cite{frutos} (F) proposes a spectral numerical method for pricing callable bonds.  The holding value function is approximated as a finite summation involving the Laguerre polynomials, and the problem is converted to a stiff system of ordinary differential equations (ODEs) for the time-dependent coefficients of the Laguerre expansion. 
\cite{bbkl} (BBKL) propose an alternative dynamic programming approach not based on PDEs in which a piecewise linear approximation is used for the value function and the exact transition probability is used to compute the expectation of the discounted piecewise linear approximation of the value function in the cases of CIR and Vasicek short rate models where the exact analytical expressions are available. All of these papers provide numerical experiments illustrating computational performance of their respective methods 
on the same example of a Swiss callable bond. This provides a natural comparison benchmark.

The present paper proposes an efficient method to evaluate callable and putable bonds under a wider class of interest rate models than any of the previous approaches, including the popular short rate diffusion models, as well as their time changed versions with jumps, including both jump-diffusion and pure jump models with state-dependent jumps. The method is based on the eigenfunction expansion of the pricing operator. We show that, under some technical conditions, the callable and putable bond pricing function has an eigenfunction expansion in eigenfunctions of the pricing operator with the expansion coefficients determined through a backward recursion. For popular short rate diffusion models, such as Cox-Ingersoll-Ross (CIR) \citep{cir}, Vasicek \citep{vasicek}, 3/2 \citep{ahngao}, we demonstrate on the test case of the Swiss bond used in the previous studies that the method is orders of magnitude faster than the alternative approaches in the literature. In addition, in contrast to the alternative approaches in the literature that have so far been limited to diffusions, the method is equally applicable to short rate jump-diffusion and pure jump models constructed from diffusion models by Bochner's subordination with a L\'{e}vy subordinator. 

The strength of the eigenfunction expansion method is that the value function is constructed globally in state and time with no need for discretization of either state or time variables. The only approximations involved in the computation scheme based on the method are the truncation of the infinite eigenfunction expansion (that, under some technical conditions, is uniformly convergent with uniformly controlled truncation error) and the numerical solution of a non-linear equation to determine the stopping boundary at each step of the backward recursion solved by the fast-converging bisection algorithm. Another strength of the method is that it  
can be seamlessly applied to both jump-diffusion and pure jump interest rate models obtained from diffusion models by subordination.
Semi-analytical methods of BW and BBKL are not suitable to handling jump-diffusion and pure jump models since no analytical solutions are available for transition probabilities and Green's functions in these models. Numerical PDE methods, such as finite-difference and finite element methods, can, in principle, be extended to handle partial integro-differential equations (PIDE) arising in jump-diffusion and pure jump models, but at substantial costs both in the theoretical complexity and structure of the schemes and in their computational implementation and computational performance. Moreover, most of the existing instances of applying numerical PIDE methods in computational finance have been limited to L\'{e}vy processes with state-independent jumps. In contrast, the eigenfunction expansion method is capable of handling models with state-dependent jumps, such as mean-reverting jumps in the interest rate. Both at the theoretical and computational level, moving from a pure diffusion model to a jump-diffusion or a pure jump model obtained from the diffusion model by the time change with respect to a L\'{e}vy subordinator amounts to no more than replacing the eigenvalues $e^{-\lambda_n t}$ of the pricing operator in the original diffusion model  with the eigenvalues $e^{-\phi(\lambda_n) t}$ of the pricing operator in the time changed model, where $\phi(\lambda)$ is the Laplace exponent of the L\'{e}vy subordinator. Remarkably, this insight goes back to the seminal work of \cite{bochner} introducing the idea of subordination (see page 370). It has been applied in probability theory (\cite{albev,chensong}) and in finance (\cite{alba,boyar,mendoza,mendoza2,lingfei2}).  The idea to use subordinated diffusions to build financial models goes back to \cite{barnlev}, who considered NIG-like Feller processes for option pricing in the pseudo-differential operator framework.

Surveys on the application of eigenfunction expansions to the valuation of European-style derivatives can be found in \cite{linet1,linethandbook}, where extensive bibliographies are given.  Applications to interest rate models and the valuation of bonds without embedded options in particular can be found in 
\cite{lewis1,davy,gorovoi,gorovoi2,boyar}. Applications of eigenfunction expansions to European-style derivatives in models with jumps constructed by time changing diffusions with L\'{e}vy subordinators can be found in \cite{alba}, \cite{boyar}, \cite{mendoza}, \cite{mendoza2},  and \cite{lingfei2}. The reference \cite{boyar} is particularly relevant to our paper, as they also consider interest rate models based on subordinated diffusions.

The rest of the paper is organized as follows.  Section 2.1 describes the general framework for the application of eigenfunction expansions to short rate diffusion models. Section 2.2 describes short rate models with jumps constructed by time-changing diffusion models with a L\'{e}vy subordinator.  Section 3 presents our eigenfunction expansion method for solving the dynamic programming backward induction for callable and putable bonds.  Section 4 presents examples of eigenfunction expansions for CIR, subordinate CIR, Vasicek, subordinate Vasicek, the 3/2 model, and the subordinate 3/2 model.  Section 5 presents numerical experiments demonstrating computational performance of the method on the test case considered in the previous callable bond literature.  Appendix contains selected proofs.

\section{Short Rate Models}

\subsection{Short Rate Diffusion Models}
\label{sec:onedim}

Let $\{X_t, t\geq 0\}$ be a one-dimensional, time-homogeneous regular (i.e. it reaches every point in $(l,r)$ with positive probability) diffusion process on the interval $I\subseteq \mathbb{R}$ with (finite or infinite) endpoints $l$ and $r$, $-\infty \leq l <r\leq \infty$.  An endpoint is unattainable if it is a natural or an entrance boundary and is attainable if it is an exit or a regular boundary (see pp.14-15 of \cite{boro} for Feller's classification of boundaries for one-dimensional diffusions).  
In this paper we assume that the endpoints are either unattainable (and, thus, not included in the state space, so the interval $I$ is open at such an endpoint) or regular and specified as instantaneously reflecting (the endpoint is included in the state space in that case, so the interval is closed at such an endpoint).
We assume that the diffusion is conservative, that is $P_t(x,I)=1$ for each $t\geq 0$ and $x\in I$, 
where $P_t(x,A)$ is the transition function from the initial state $x$ to the Borel set $A\subseteq I$ in time $t$.
Thus the process $X$ has infinite lifetime.  We assume that the volatility $\sigma(x)$ of $X$ is positive and continuous on the open interval $(l,r)$ and the drift $\mu(x)$ is continuous on $(l,r)$.

We further assume that the instantaneous interest rate (the short rate) $r_t$ at time $t$ is a function of the state variable $X_t$ and is given by $r_t=r(X_t)$.  We assume that $r(x)$ is continuous on $(l,r)$. The continuity assumptions for $\sigma,$ $\mu$ and $r$ are not necessary, but simplify exposition in what follows.  
Consider the family of {\em pricing operators} (mathematically, {\em Feynman-Kac (FK) operators}) $\{\mathcal{P}_t^r, t\geq 0 \}$ defined by 
$$
\mathcal{P}_t^r f(x):={\mathbb E}_x\left[e^{-\int_0^t r(X_u) \, \mathrm{d}u } f(X_t) \right],  
$$
where ${\mathbb E}_x$ is the expectation operator with respect to the probability measure ${\mathbb P}_x$ of the process $X$ starting at $x\in I$ at time zero. 
Since in this paper we are interested in pricing, we always work with risk-neutral probabilities chosen by the market.
The pricing operator ${\cal P}_t^r$ maps future payoff functions of the future state at time $t$ into present value functions of the present state at time zero by discounting from time $t$ to time zero and taking the expectation conditional on the current state at time zero. Under our assumptions, these operators form a strongly continuous semigroup on the Banach space $C_b(I)$ of bounded continuous functions on $I$ with the supremum norm (see, e.g., \citet{boro}). 
The infinitesimal generator $\mathcal{G}^r$ of this semigroup acts on twice continuously differentiable functions on $I$ with compact supports by the second-order differential operator (the {\em Sturm-Liouville operator}):
$$
\mathcal{G}^rf(x) = \frac{1}{2}\sigma^2(x)f''(x)+\mu(x)f'(x)-r(x)f(x).
$$  

Define $s(x)$ and $m(x)$ to be the {\em scale} and {\em speed densities} of the diffusion process $X$:
$$
s(x) = \exp\left\{-\int_{x_0}^x \frac{2\mu(y)}{\sigma^2(y)}\, \mathrm{d}y \right\}, \quad  m(x) = \frac{2}{\sigma^2(x)s(x)},
$$ 
where $x_0 \in (l,r)$ is an arbitrary point (see \cite{karlin}, \cite{boro} for discussions of the scale function and the speed measure of the one-dimensional diffusion).  The infinitesimal generator can then be re-written in the formally self-adjoint form:
$$
\mathcal{G}^r f(x) =\frac{1}{m(x)}\left(\frac{f'(x)}{s(x)}\right)'-r(x)f(x).
$$ 
Under our assumptions, the generator ${\cal G}^r$ and the FK semigroup $\{{\cal P}_t^r,t\geq 0\}$  can be extended to a self-adjoint operator and the symmetric strongly-continuous semigroup in the Hilbert space $L^2(I,m)$ of functions on $I$ square-integrable with the speed measure $m(dx)=m(x)dx$ and endowed with the inner product 
$$
(f,g)=\int_I f(x)g(x) m(x) \, \mathrm{d}x
$$
and norm $\|f\|=\sqrt{(f,f)}$.
Thus, the Spectral Theorem for self-adjoint operators in Hilbert spaces can now be applied to write down the spectral decomposition of the generator and the semigroup.
The spectral representation for one-dimensional diffusions goes back to the classical work of \citet{mckean}.  We refer the reader to \citet{linet1,linethandbook} for surveys of applications in finance. 

In this paper we further assume that the negative of the infinitesimal generator $-{\cal G}^r$ has a purely discrete spectrum in $L^2(I,m)$ bounded from below.  
Sufficient conditions for the purely discrete spectrum in terms of the behavior of the functions $\sigma$, $\mu$ and $r$ near the boundaries $l$ and $r$ are given in 
 \citet{linet1,linethandbook}.  When the spectrum of $-{\cal G}^r$ is purely discrete and bounded from below, the FK semigroup has the eigenfunction expansion of the form:
\begin{equation} \label{eigenexpansion}
\mathcal{P}_t^r f(x)=\sum_{n=0}^\infty  f_n e^{-\lambda_n t} \varphi_n(x), \quad f_n=(f,\varphi_n),
\end{equation}
for any $f\in L^2(I,m)$, where $\{\lambda_n\}_{n=0}^\infty$ such that $\lambda_0 < \lambda_1 < \cdots$, $\lim_{n \uparrow \infty} \lambda_n = \infty$, are the eigenvalues of $-\mathcal{G}^r$ and $\varphi_n$ are the corresponding eigenfunctions normalized so that $\|\varphi_n\|^2 = 1$ (for future convenience we index the eigenvalues and eigenfunctions starting from $n=0$ rather than $n=1$).  The eigenfunctions form a complete orthonormal basis in $L^2(I,m)$.
We also assume that the eigenvalues satisfy the condition:
\begin{equation}\label{traceclass}
\sum_{n=0}^\infty e^{-\lambda_n t} <\infty
\end{equation}
for all $t>0$, so that the FK semigroup is {\cal trace class} (see Section 7.2 in \citet{davies}). We recall that the semigroup of a one-dimensional diffusion has a symmetric density $p_t(x,y)$ with respect to the speed measure $m(dx)=m(x)dx$ that is a continuous function in $t$, $x$ and $y$
(see p.149 in \citet{itomckean} or p.13 of \citet{boro}).  Hence, according to Theorem 7.2.5 in \citet{davies}, the eigenfunctions $\varphi_n(x)$ are continuous functions with the global estimate $|\varphi_n(x)|\leq e^{\lambda_nt/2}\sqrt{p_t(x,x)}$ for all $t>0$, and the density $p_t(x,y)$ has an eigenfunction expansion for all $t>0$
\begin{equation}
\label{transitioneig1}
p_t(x,y)=\sum_{n=0}^\infty e^{-\lambda_n t}\varphi_n(x)\varphi_n(y)
\end{equation}
that converges uniformly in $x$ and $y$ on compacts. This ensures that, in addition to the $L^2$ convergence, the eigenfunction expansion \eqref{eigenexpansion} converges uniformly in $x$ on compacts for all $f\in L^2(I,m)$ and $t>0$. This follows from the Cauchy-Schwartz bound for the expansion coefficients $|f_n|\leq \|f\|$, the eigenfunction estimate, and the trace class condition \eqref{traceclass}.    

Since we are interested in bond pricing in this paper, we assume that the constant payoffs are in the Hilbert space $L^2(I,m)$, i.e. $1\in L^2(I,m)$. This is equivalent to assuming that the speed measure $m$ is a finite measure on $I$, $m(I)<\infty$. In that case, the speed density can be normalized to one to be a probability density and, thus, serves as the steady state density of the underlying process $X$. Then the present value at time zero of 
a zero-coupon bond with unit face value and maturity $t\geq 0$ when the underlying process is in state $x$, $X_0=x$, has the eigenfunction expansion 
 given by: 
\begin{equation} \label{bondprice}
P(t,x) = {\mathbb E}_x\left[e^{-\int_0^t r(X_u) \, \mathrm{d}u } \right] = \sum_{n=0}^\infty p_n e^{-\lambda_n t} \varphi_n(x)
\end{equation}
with the expansion coefficients $p_n = (1,\varphi_n)$.  Under our assumptions, the expansion converges uniformly in $x$ on compacts for all $t>0$.

Virtually all popular short rate models in the financial economics literature fit into the general framework described above, including
Cox-Ingersoll-Ross (CIR) model \citep{cir}, Vasicek model \citep{vasicek}, the 3/2 model \citep{ahngao},
Black's model of interest rates as options \citep{gorovoi}, and
the quadratic model \citep{beagle,leip}. Note that we have not made the assumption that $r(x)$ is non-negative to accommodate the Vasicek model. If we make that assumption, then ${\cal G}^r$ is positive semi-definite (so that $-{\cal G}^r$ is negative semi-definite), the FK semigroup $({\cal P}_t^r)_{t\geq 0}$ is a contraction semigroup on $L^2(I,m)$, and $\lambda_0\geq 0$. To accommodate the Vasicek model, we made a weaker assumption that the spectrum of $-{\cal G}^r$ is bounded from below, rather than non-negative.

\subsection{Short Rate Models with Jumps Constructed by Subordination}

\label{sec:subdiff}

A {\em subordinator} $\{T_t, t\geq 0 \}$ is a nondecreasing L\'{e}vy process with the Laplace transform given by the L\'{e}vy-Kchintchine formula 
$$
E\left[e^{-\lambda T_t} \right] = e^{-t \phi(\lambda)}, \quad \phi(\lambda) = \gamma \lambda + \int_0^{\infty} (1-e^{-\lambda s}) \nu(\mathrm{d}s),\quad \lambda \geq 0$$
with the Laplace exponent $\phi(\lambda)$, 
with nonnegative drift $\gamma \geq 0$, and L\'{e}vy measure $\nu(\mathrm{d}s)$ satisfying the integrability condition $\int_0^{\infty} (s\wedge 1) \nu (\mathrm{d} s) < \infty$.  
For any set $A\subset \mathbb{R}$ bounded away from zero, jumps of sizes in $A$ arrive according to a Poisson process with the arrival rate $\nu(A)$. If $\nu$ is a finite measure on $(0,\infty)$, the subordinator is a compound Poisson process plus drift at the rate $\gamma$. If $\nu(0,\infty)=\infty$, the subordinator is a jump process with infinite activity and drift $\gamma$. If $\gamma=0$, it is a pure jump process. 
Examples of subordinators important in applications include compound Poisson processes with exponential or gamma distributed jump sizes, inverse Gaussian (IG) subordinators \citep{barndorff}, and gamma subordinators \citep{madan}, with the latter two having infinite activity.  These examples are special cases of subordinators with L\'{e}vy measures of the form $\nu(\mathrm{d}s) = C s^{-p-1} e^{-\eta s} \mathrm{d}s$ with $C>0$, $\eta >0$, and $p<1$.  The case with $p\in (0,1)$ are the tempered stable subordinators (the limiting cases with $\eta=0$ are stable subordinators). The special case with $p=1/2$ is the IG subordinator. The limiting case with $p=0$ is the Gamma subordinator. 
The Laplace exponent is given by:
\begin{equation}
\phi(\lambda) = \begin{cases}
\gamma \lambda - C\Gamma(-p) \left[(\lambda+\eta)^p - \eta^p \right], &p\neq 0\\
\gamma \lambda + C \ln (1+\lambda/\eta), & p=0
\end{cases}.
\label{temper}
\end{equation}  
As an example, the L\'{e}vy measure and Laplace exponent for an IG subordinator parameterized with $\mu$ and $\nu$, the mean and variance of an IG process at time one, $t=1$, are given by:
\begin{equation}
\nu(\mathrm{d}s) = \mu \sqrt{\frac{\mu}{2\pi\nu}}s^{-\frac{3}{2}}\exp\left\{-\frac{\mu}{2\nu}s\right\} \mathrm{d}s,\quad \phi(\lambda) = \gamma \lambda+ \frac{\mu^2}{\nu}\left( \sqrt{1+2\frac{\nu}{\mu} \lambda}-1 \right). \label{nugauss} 
\end{equation}
Further mathematical details on subordinators can be found in \citet{schill} and on financial applications in \citet{cont}. 

Since subordinators are non-negative, non-decreasing processes, they can be used as stochastic time changes to time change other processes. This procedure is known as {\em Bochner's subordination} and goes back to \citet{bochner, bochner2}. In particular, time changing a Markov process with a subordinator yields another Markov process whose semigroup and infinitesimal generator are given by Phillips' theorem (\cite{phillip}, Theorem 32.1 in \cite{sato}, Chapter 12 in \citet{schill}). 
For recent financial applications see  \cite{mendoza}, \cite{mendoza2}, and \citet{lingfei2}.

In particular, we can construct new short rate models with jumps from diffusion short rate models as follows.
Let $X$ be an ergodic one-dimensional diffusion on $I$ with volatility $\sigma(x)$ and drift $\mu(x)$ and with the stationary density given by the (normalized) speed density $m$  as described in the previous section. 
Let $r(x)$ be the function defining the diffusion short rate model as in section 2.1 and $\{{\cal P}_t^r,t\geq 0\}$ the corresponding FK semigroup.
Then a new short rate model with jumps is obtained by subordinating the original diffusion short rate model with respect to a given subordinator ${\cal T}$ with the Laplace exponent $\phi(\lambda)$ by defining a new semigroup (the superscript $\phi$ signifies that the subordination is performed with respect to the subordinator with the Laplace exponent $\phi$):
\begin{equation} \label{suboperator}
\mathcal{P}_t^{r,\phi} f(x):={\mathbb E}_x\left[e^{-\int_0^t r^\phi(Y_u) \, \mathrm{d}u } f(Y_t) \right], \quad t\geq 0,
\end{equation}
where $Y_t$ is a new Markov jump-diffusion process on $I$ with the infinitesimal generator ${\cal G}^\phi$
acting on twice-differentiable functions with compact supports as an integro-differential operator (see  \citet{mendoza} and \citet{mendoza2}):
$$
\mathcal{G}^\phi f(x) = \frac{1}{2}\gamma \sigma^2(x) f''(x)+\mu^\phi(x)f'(x)
+\int_{(l,r)} \left( f(y)-f(x)-\mathbf{1}_{\{|y-x|\leq 1 \}} (y-x) \frac{\mathrm{d}f}{\mathrm{d}x}(x) \right) \pi^\phi(x,y) \, \mathrm{d}y,
$$
where the drift with respect to the truncation function $\mathbf{1}_{\{|y-x|\leq 1 \}}$ is
$$
\mu^\phi(x) = \gamma \mu(x)+\int_{(0,\infty)} \left( \int_{\{y \in (l,r): |y-x| \leq 1 \}} (y-x) p_s(x,y)  \, \mathrm{d}y \right) \nu(\mathrm{d}s),
$$
and $\pi^\phi(x,y)$ is the symmetric state-dependent L\'{e}vy density:
$$
\pi^\phi(x,y)=\int_{(0,\infty)} p_s(x,y) \nu(\mathrm{d}s),
$$
where $p_s(x,y)$ is the density of the original FK semigroup $({\cal P}_t^r)_{t\geq 0}$ of the pure diffusion short rate model, and $\gamma$  and $\nu$ are the drift and the L\'{e}vy measure of the subordinator. When $\gamma>0$, $Y$ is a jump-diffusion. When $\gamma=0$, $Y$ is a pure jump process. $Y$ has the same steady state density $m(x)$ as the original diffusion $X$.
The short rate in \eqref{suboperator} is the function of the state variable $Y_t$ so that $r_t=r^\phi(Y_t)$ with the function $r^\phi(x)$ given by: 
$$
r^\phi(x)=\gamma r(x)+\int_{(0,\infty)}(1-P(s,x))\nu(\mathrm{d}s),
$$
where $P(s,x)$ is the price \eqref{bondprice} of the $s$-maturity zero-coupon bond at time zero when the state of the underlying diffusion $X_0=x$. The form of the generator ${\cal G}^\phi$ of $Y$ and the function $r^\phi$ follow from Phillips' theorem that characterizes the subordinate semigroup and its infinitesimal generator. Here we subordinate the pricing (FK) semigroup of the diffusion process $X$ with the discount rate $r_t=r(X_t)$ with respect to a given subordinator to obtain a new semigroup interpreted as the pricing (FK) semigroup of a new short rate model $r_t=r^\phi(Y_t)$ driven by the state variable $Y_t$ following a Markov process with jumps.

\begin{remark} Mathematically, the subordination of the FK semigroup can be interpreted as follows. First formulate the FK semigroup ${\cal P}^r$ of the original conservative diffusion $X$ with the discount rate $r(x)$ as the transition semigroup of the diffusion $\hat{X}$ with killing at the rate $r(x)$ (cf. Section II.4 on pp.27-28 in \citet{boro} for the connection between discounting and killing). Then construct a new process $\hat{X}^\phi_t:=\hat{X}_{{\cal T}_t}$ by time changing $\hat{X}$ with the subordinator ${\cal T}$. 
Use Phillips' theorem to write down its infinitesimal generator and, thus, its local characteristics (diffusion, drift, state-dependent L\'{e}vy measure, and state-dependent killing rate $r^\phi(x)$).
Then formulate the transition semigroup of the process $\hat{X}^\phi$ as the FK semigroup of a conservative process $Y$ with the generator given above and with the discount rate $r^\phi(x)$ given above. The formulation of the application of Phillips' theorem to this situation is given in \cite{mendoza} and we do not repeat it here to save space.
\end{remark}

The subordinate FK semigroup $\{{\cal P}_t^{r,\phi},t\geq 0\}$ is also a symmetric strongly continuous semigroup of operators on $L^2(I,m)$ \citep{chen}, and, under the assumptions we have made about the semigroup $\{{\cal P}_t^{r},t\geq 0\}$ in the previous section, it possesses an eigenfunction expansion in the {\em same} eigenfunctions $\varphi_n(x)$ with $\lambda_n$ in Eq.\eqref{eigenexpansion} replaced with $\lambda_n^\phi:=\phi(\lambda_n)$, where $\phi(\lambda)$ is the Laplace exponent of the subordinator:
\begin{equation} \label{subeigenexpansion}
\mathcal{P}_t^{r,\phi} f(x)=\sum_{n=0}^\infty  f_n e^{-\phi(\lambda_n) t} \varphi_n(x), \; f_n=(f,\varphi_n),
\end{equation}
for any $f\in L^2(I,m)$ and $t>0$.
For mathematical details on subordination of semigroups of operators and Markov processes see the excellent exposition in \citet{schill}, and for recent financial applications see \citet{mendoza}, \citet{mendoza2}, and \citet{lingfei2}.
If we further assume that the Laplace exponent $\phi$ of the subordinator is such that it satisfies  the condition
\begin{equation} \label{subuniform}
\sum_{n=0}^\infty e^{-\phi(\lambda_n) t}<\infty
\end{equation}
for all $t>0$, then the semigroup $\{{\cal P}_t^{r,\phi},t\geq 0\}$ is also trace class. 
If we further assume that the eigenfunctions of the original pure diffusion FK semigroup have a bound independent of $n$ on each compact interval $K=[a,b]\subset (l,r)$, i.e, $|\varphi_n(x)|\leq C_K$ for all $n$, where the constants $C_K$ may depend on the interval $K$ but are independent of $n$, then  these assumptions ensures that, in addition to the $L^2$ convergence, the eigenfunction expansion of the subordinate semigroup \eqref{subeigenexpansion} converges uniformly in $x$ on compacts for all $f\in L^2(I,m)$ and $t>0$ and that the subordinate semigroup $\{{\cal P}_t^{r,\phi},t\geq 0\}$ also has a continuous density with respect to $m(x)dx$ with the eigenfunction expansion for all $t>0$
\begin{equation} \label{transitioneig2}
p_t^\phi(x,y)=\sum_{n=0}^\infty e^{-\phi(\lambda_n)t}\varphi_n(x)\varphi_n(y)
\end{equation}
uniformly convergent on compacts in $x$ and $y$.

For the present value at time zero of a zero-coupon bond with unit face value and maturity $t\geq 0$ when the underlying state variable has initial value of $Y_0=x$,  we then obtain the eigenfunction expansion given by: 
\begin{equation} \label{subzerobond}
P(t,x) = {\mathbb E}_x\left[e^{-\int_0^t r^\phi(Y_u) \, \mathrm{d}u } \right] = \sum_{n=0}^\infty p_n e^{-\phi(\lambda_n)t} \varphi_n(x)
\end{equation}
with the expansion coefficients $p_n = (1,\varphi_n)$.  Under the assumption \eqref{subuniform} on the growth $\phi(\lambda)$ and the bound on eigenfunctions independent of $n$, the expansion converges uniformly in $x$ on compacts for all $t>0$.

Using the subordination approach, we can extend all the diffusion short rate models popular in financial economics to jump-diffusion and pure jump models, in particular constructing subordinate CIR (SubCIR), subordinate Vasicek (SubVasicek), etc. 
Subordinate models allow for jumps in the interest rate dynamics. Moreover, if the diffusion process is mean-reverting, the subordinate process will have jumps that are also mean-reverting (see \citet{lingfei2} for the proof in the subordinate Ornstein-Uhlenbeck context). While adding jumps improves the model's realism and flexibility, remarkably, the analytical and computational framework remains entirely unchanged, as the only modification required in the eigenfunction expansion is the replacement of $\lambda_n$ in Eq.\eqref{eigenexpansion} with $\lambda^\phi_n=\phi(\lambda_n)$ in Eq.\eqref{subeigenexpansion}.

\begin{remark}{\em Matching the initial yield curve.}
The time-homogeneous short rate models discussed in sections 2.1 and 2.2 can be extended to match any initial term structure of interest rates
as proposed by \cite{brigo2} and commonly done in fixed income market practice by adding a deterministic function of time to the short rate process in the extended diffusion and the subordinated diffusion models, respectively.  The function can then be chosen so that the initial zero-coupon bond prices of all maturities in the extended model match the zero-coupon prices consistent with the given initial term structure of interest rates (given yield curve). The callable and putable bond pricing developed in this paper can then be immediately extended to this class of models. To simplify notation, we do not explicitly consider this extension in what follows and assume the short rate model is time homogeneous.  
\end{remark}

\section{The Eigenfunction Expansion Method for Callable and Putable Bonds}

The call option allows the bond issuer to buy the bond back from the bond holder (call the bond) for pre-specified call prices at some pre-specified times prior to maturity. This allows the bond issuer to refinance the bond if interest rates decline.  The put option allows the bond holder to sell (put) the bond back to the bond issuer for pre-specified put prices at some pre-specified times prior to maturity. This allows the bond holder to re-invest the proceeds into a bond with higher coupon if interest rates rise.  We assume that the bond principal is equal to one dollar and the bond pays coupons of $C$ dollars on the coupon dates $t_i$, $i=1,...,k$.  Let the bond issue date and maturity date be $t_0=0$ and $t_k=T$, respectively.  After some initial protection period from $t_0$ until $t_{k^{*}}$, the call and put options can be exercised at the subsequent coupon dates $t_i$, $i=k^{*},...,k-1$, prior to maturity $t_k$.  We also assume that there are {\em notice periods} of lengths $\delta$ so that the option exercise decision to exercise at the coupon date $t_i$ has to be made at an earlier time $\tau_i=t_i-\delta$ so that the adequate advance notice of duration $\delta$ can be given to the other party.  Denote the call and put prices at times $t_i$ as $K_i^c$ and $K_i^p$, respectively.  It is assumed that $K_i^c > K_i^p$.  This sets up an optimal stopping game in discrete time with finite horizon where one player (the bond holder) chooses a stopping time to maximize the bond value and the other player (the bond issuer) chooses a stopping time to minimize the bond value.  

We assume that the short rate is $r_t=r^\phi(Y_t)$, where $Y_t$ is a subordinate diffusion, as discussed in section 2. We note that the pure diffusion model can be viewed a special case with the trivial time change ${\cal T}_t=t$ with $\phi(\lambda)=\lambda$, $\gamma=1$ and $\nu\equiv 0$. We assume that all the assumptions made in section 2 are in force. Then the pricing operator ${\cal P}_t^{r,\phi}$ has the eigenfunction expansion \eqref{subzerobond}. To simplify notation, we drop the superscripts $r$ and $\phi$ and simply write ${\cal P}_t$ for the pricing operator ${\cal P}_t^{r,\phi}$ in what follows.

Let $V(t,x)$ be the value of the bond at time $t\in [0,T]$ when the underlying state is $x$.
Since the decisions to exercise the call and put options are made at times $\tau_i=t_i-\delta$ prior to the coupon dates, the present values at $\tau_i$ of the call and put prices at time $t_i$ must be compared to the holding (continuation) values of the bond at time $\tau_i$.  The  discounted value of the call price at time $\tau_i$ is $K_i^c P(\delta,x)$, where $P(\delta,x)$ is the value of the zero-coupon bond with time to maturity equal to the notice period $\delta$ and unit face value when the state variable is in the state $x$ at time $\tau_i$.  The expected discounted value of the put price at time $\tau_i$ is $K_i^p P(\delta,x)$. 
At maturity the bond value is equal to its principal plus the last coupon, $V(t_k,x)=1+C$. 
Let $V^k(x):=V(t_k,x)$ and, for $i\leq k-1$, $V^i(x):=V(\tau_i,x)$ denote the bond's value at time $\tau_i$  and $C^i(x)$ denote the bond's holding (continuation) value at time $\tau_i$ {\em ex-coupon} at time $t_i$, $i=k^*,...,k-1$ (the assumption is that the next coupon $C$ is always paid at time $t_i$, whether or not decisions are made to exercise any of the two options at time $\tau_i$). Let $V^0(x):=V(0,x)$ denote the value of the bond at the time of issue.

Assuming both players behave rationally and maximize the value of their assets and minimize the value of their liabilities,
in state $x$ the bond issuer will exercise the call option at time $\tau_i$ if $K_i^cP(\delta,x)\leq C^i(x)$, the bond holder will exercise the put option at time $\tau_i$ if $K_i^pP(\delta,x)\geq C^i(x)$, and there will be no exercise of either option at time $\tau_i$ if $K_i^pP(\delta,x)< C^i(x)< K_i^cP(\delta,x)$.  By the assumption that $K_i^c > K_i^p$ the simultaneous exercise of both options is never optimal. Further set $h_i:=\tau_{i+1}-\tau_i$, $i=k^*,...,k-2,$ and $h_{k-1}:=t_k-\tau_{k-1}$.
Then the 
value of the bond with call and put options satisfies the following Bellman's dynamic programming backward induction (see also \cite{bbkl}):
\begin{equation} \label{vk} 
V^{k}(x) = (1+C), 
\end{equation}
\begin{equation} \label{continuation}
C^i(x)={\cal P}_{h_i}V^{i+1}(x),\quad \;k^* \leq i \leq k-1, 
\end{equation}
\begin{equation}\label{vtau} 
V^i(x) = \max \left\{K_i^p P(\delta,x), \min \left\{K_i^c P(\delta,x), C^i(x) \right\} \right\} +CP(\delta,x), \;k^* \leq i \leq k-1, 
\end{equation}
\begin{equation} 
\label{v0} 
V^0(x) = {\cal P}_{\tau_{k^*}}V^{k^*}(x)+C \sum_{i=1}^{k^{*}-1} P(t_i,x).
\end{equation}

Assuming that for each $i=k^*,...,k-1$ each of the two equations
\begin{equation} \label{breakeqn}
K_i^cP(\delta,x)=C^i(x),\quad K_i^p P(\delta,x)=C^i(x)
\end{equation}
has at most one solution in $I$ denoted by $x_i^c$ and $x_i^p$, respectively, setting $x_i^c:=l$ if the first equation has no solution in $I$, setting $x_i^p:=r$ if the second equation has no solution in $I$, and  noting that $x_i^c<x_i^p$, the backward induction \eqref{vtau} can be re-written in the form:
\begin{equation} 
V^i(x) = K_i^c P(\delta,x){\bf 1}_{\{x< x_i^c\}} + K_i^p P(\delta,x){\bf 1}_{\{x>x_i^p\}}+ C^i(x){\bf 1}_{\{x_i^c\leq x\leq x_i^p\}}  + C P(\delta,x),
\end{equation}
for all $k^* \leq i \leq k-1$ and $x\in I$, where ${\bf 1}_{\{\cdot\}}$ denotes the indicator function, and ${\bf 1}_{\{x<x_i^c\}}\equiv 0$ if $x_i^c=l$ and ${\bf 1}_{\{x>x_i^p\}}\equiv 0$ if $x_i^p=r$ by convention.

This backward induction can be solved by a variety of computational methods in the literature, as discussed in the introduction and in section 5, based on the different methods to approximate the pricing operator ${\cal P}_t$ appearing in Eqs.\eqref{continuation} and \eqref{v0}.
In this paper we follow the 
 approach 
 based on representing the pricing operator ${\cal P}_t$ by its eigenfunction expansion.  Our main result is the following theorem that summarizes our eigenfunction expansion method for the valuation of callable and putable bonds.
\begin{thm}
\label{section3theorem}
Suppose that $m$ is a finite measure on $I$, $m(I)<\infty$. 

(i)  The value function $V^0(x)$ and the value functions $V^i(x)$ and the continuation value functions $C^i(x)$ are in $L^2(I,m)$ for all $i=k^*,...,k-1$.

(ii)  The continuation value functions have the following eigenfunction expansions:
\begin{equation}
\label{vholdeig1}
C^i(x) = \sum_{n=0}^\infty c_n^{i+1} e^{-\lambda_n h_i} \varphi_n(x),\quad i=k^*,...,k-1.
\end{equation}
The value function at the time of the bond issue has the following eigenfunction expansion:
\begin{equation} \label{v0thm}
V^0(x) = \sum_{n=0}^\infty c_n^{k^*} e^{-\lambda_n \tau_{k^*}} \varphi_n(x)+ C \sum_{i=1}^{k^*-1} P(t_i,x).
\end{equation}

(iii) The eigenfunction expansions \eqref{vholdeig1} and \eqref{v0thm} converge uniformly in $x$ on compacts in $I$.

(iv) If each of the two equations in \eqref{breakeqn} has at most one solution in $I$, then the eigenfunction expansion coefficients in \eqref{vholdeig1} and \eqref{v0thm} satisfy the following backward recursion:
$$
c^k_n=(1+C)p_n, \; n=0,1,2,...,
$$ 
where $p_n=(1,\varphi_n)$ are the expansion coefficients of the unit payoff appearing in the eigenfunction expansion of the zero-coupon bond \eqref{subzerobond},
and for each $i=k^*,...,k-1,$
\begin{equation} \label{coeffrecurthm}
c^i_n =   K_i^c p_n(l,x_i^c) + \sum_{m=0}^\infty c_m^{i+1}e^{-\lambda_m h_i} \pi_{m,n}(x_i^c,x_i^p)              +   K_i^p p_n(x_i^p,r)  + C p_n e^{-\lambda_n \delta}, 
\end{equation}
for $n=0,1,2,...$, 
where $x_i^c$ and $x_i^p$ are as previously defined, and for $l\leq x < y\leq r$ we introduced the following notation
\begin{equation} \label{pieqn}
\pi_{m,n}(x,y):=({\bf 1}_{(x,y)}\varphi_m,\varphi_n)=\int_x^y \varphi_m(z)\varphi_n(z)m(z) \, \mathrm{d}z,
\end{equation}
\begin{equation} \label{pneqn}
p_n(x,y):=( {\bf 1}_{(x,y)}P(\delta), \varphi_n)=\int_{x}^{y} P(\delta,z) \varphi_n(z) m(z) \, \mathrm{d}z,
\end{equation}
where ${\bf 1}_{(x,y)}={\bf 1}_{(x,y)}(z)$ is the indicator function of the interval $(x,y)$ and $P(t)=P(t,x)$ is the value function of the zero-coupon bond with time to maturity $t\geq 0$ and unit face value when the underlying state variable is $x$. 
\end{thm}
\begin{proof}
See the Appendix.
\end{proof}

Theorem 3.1 reduces the solution of the backward induction for callable and putable bonds to the recursion for the expansion coefficients \eqref{coeffrecurthm}, together with finding the roots of equations \eqref{breakeqn} by a numerical root finding algorithm, such as bisection. The continuation value $C^i(x)$ on the right hand side of the equations \eqref{breakeqn} is given by the eigenfunction expansion \eqref{vholdeig1} with the coefficients determined on the previous step of the recursion. The eigenfunction expansion can be truncated at a finite level, and the truncation error is uniformly controlled due to the uniform convergence of the expansions when condition \eqref{subuniform} is satisfied.

\begin{remark}
When the state process is an affine diffusion, such as CIR or Vasicek, the zero-coupon bond value function has the exponential-affine form in the state variable 
\begin{equation} \label{zerobondexp}
P(t,x)=A(t)e^{-B(t) x},
\end{equation} 
and the integral in Eq.\eqref{pneqn} can be written as 
\begin{equation} \label{pnzerobond}
p_n(x,y)=\int_{x}^{y} P(t,z) \varphi_n(z) m(z)\mathrm{d}z = A(t) \int_{x}^{y} e^{-B(t)z} \varphi_n(z) m(z)\mathrm{d}z 
\end{equation}
and in some cases can be calculated in closed form.
Generally, it can be calculated using the eigenfunction expansion for the zero-coupon bond value function:
\begin{equation} \label{generalzerobond}
({\bf 1}_{(x,y)}P(t),\varphi_n)=\sum_{m=0}^\infty p_m e^{-\phi(\lambda_m)t}\pi_{m,n}(x,y),
\end{equation}
where $\pi_{m,n}(x,y)$ are defined in Eq.\eqref{pieqn}.\\
\end{remark}

\begin{remark} If the bond has only the call option and no put option, then the game reduces to the optimal stopping problem for the bond issuer. In that case in Eq.\eqref{coeffrecurthm} the term with $K_i^p p_n(x_p^i,r)$ is absent and $\pi_{m,n}(x_i^c,x_i^p)$ is replaced with $\pi_{m,n}(x_i^c,r)$ for all $i$. Similarly, if the bond has only the put option and no call option, the game reduces to the optimal stopping problem for the bond holder and  in Eq.\eqref{coeffrecurthm} the term with $K_i^c p_n(l,x_c^i)$ is absent and $\pi_{m,n}(x_i^c,x_i^p)$ is replaced with $\pi_{m,n}(l,x_i^p)$ for all $i$. 
\end{remark}

The condition that each of the two equations in \eqref{breakeqn} has at most one solution in $I$ generally needs to be checked case by case.  For CIR, Vasicek, and 3/2 models considered in section 4  the condition can be verified by the following proposition.  For CIR and Vasicek, \eqref{comparison} below follows from  Theorem 1.1 in \citet{ikeda} while \eqref{comparison} holds for the 3/2 model since the diffusion process $X(t)$ in the 3/2 model can be written as $X(t)=1/Y(t)$, where $Y(t)$ is a CIR process satisfying the Feller condition.

\begin{propo}
\label{atmost}
Suppose that the discount rate $r(x)$ is a non-decreasing function.  Let $(\Omega,\mathcal{F}, \mathbb {P})$ be a complete probability space with right continuous increasing family $(\mathcal{F}_t)_{t\geq 0}$ of sub $\sigma$-fields of $\mathcal{F}$ each containing $P$-null sets and let $B_t$ be a one-dimensional $\mathcal{F}_t$-Brownian motion.  Let $\sigma(x)$ and $\mu(x)$ be continuous.  Let $X_1(t)$ and $X_2(t)$ be processes started at different initial states such that $X_1(0) < X_2(0)$ and $$X_i(t)=X_i(0)+\int_0^t \sigma(X_i(s)) \,\mathrm{d}B(s)+\int_0^t \mu(X_i(s)) \, \mathrm{d}s, \quad i=1,2.$$
If we have that
\begin{equation} \label{comparison}
\mathbb{P}[X_1(t) \leq X_2(t) \textrm{ for all } t\geq 0] =1,
\end{equation}
then each of the two equations in \eqref{breakeqn} has at most one solution in $I$.
\end{propo}
\begin{proof}
See the Appendix.
\end{proof}

\begin{remark} We emphasize that uniqueness of roots of \eqref{breakeqn} is {\em not} a requirement for our method to work. In fact, it is one of the strengths of our approach that it can handle just as easily more general cases with multiple break-even points and, hence, early exercise regions that are not necessarily one-sided and, in general, can be unions of multiple intervals. Proving the uniqueness of the break-even point provides a convenience for numerical implementation, as we can stop after finding a single root. In general, without the proof of uniqueness, a more thorough numerical investigation of the functions is required in each case to either establish uniqueness or determine multiple roots. While we have been able to prove in Proposition 3.2 uniqueness for pure diffusion short rate models, we have been unable to extend the proof to the case of subordinate diffusions, as it is based on classical SDE comparison results that to the best of our knowledge are not available in general for subordinate diffusions. Nevertheless, in our extensive numerical experimentation in all cases of subordinate diffusions we have considered, we have observed similar behavior of functions in \eqref{breakeqn} that lead to unique solutions. We thus conjecture that uniqueness also holds for subordinate diffusions, perhaps subject to some condition on the subordinator. 
\end{remark}

\section{Examples} 
\label{sec:examples}

\subsection{CIR and SubCIR} \label{seccir}

In the Cox-Ingersoll-Ross (CIR) model \citep{cir}, the short rate follows the CIR diffusion with drift $\mu(x)=\kappa(\theta-x)$ and volatility $\sigma(x)=\sigma \sqrt{x}$ where
$\kappa>0$, $\theta>0$, and $\sigma>0$ are the rate of mean reversion, the long run mean and volatility, respectively.
In this case $r(x)=x$.
When Feller's condition $2\kappa \theta / \sigma^2 \geq 1$ is satisfied, the origin is an unattainable entrance boundary and infinity is an unattainable natural boundary.  In this case $I=(0,\infty)$.  When Feller's condition is not satisfied, the origin is an attainable regular boundary and is specified as instantaneously reflecting. In this case $I=[0,\infty)$.  The CIR speed density reads $m(x)=\frac{2}{\sigma^2}x^{b-1}e^{-\frac{2\kappa x}{\sigma^2}}$.

The celebrated CIR zero-coupon bond pricing formula is:
\begin{equation} \label{cirzerobond} 
P(t,x)=A(t)e^{-B(t)x},
\end{equation} 
where 
\begin{equation*} 
A(t)=\left(\frac{2\gamma e^{(\kappa+\gamma)t/2}}{(\gamma+\kappa)(e^{\gamma t}-1)+2\gamma}  \right)^b,\quad
B(t)=\frac{2(e^{\gamma t}-1)}{(\gamma+\kappa)(e^{\gamma t}-1)+2\gamma}, 
\end{equation*}
\begin{equation} \label{gamb} 
\gamma=\sqrt{\kappa^2+2\sigma^2},\; b=\frac{2\kappa \theta}{\sigma^2}. 
\end{equation}

The eigenfunction expansion \eqref{bondprice} of the CIR zero-coupon bond pricing function is given in
\cite{davy}. In this case the eigenfunctions, eigenvalues, and the expansion coefficients for the unit payoff are:
$$
\lambda_n =\gamma n+\frac{b}{2}(\gamma-\kappa),
$$
$$
\varphi_n(x)=N_n e^{((\kappa-\gamma)x)/\sigma^2}L_n^{(b-1)}\left(\frac{2\gamma x}{\sigma^2}\right),\quad N_n=\sqrt{\frac{\sigma^2 n!}{2\Gamma(b+n)}}\left(\frac{2\gamma}{\sigma^2}\right)^{b/2}
$$
\begin{equation*} 
\label{cn} 
p_n =(1,\varphi_n)= \frac{2N_n \Gamma(b+n)}{\sigma^2 n!}\left(\frac{\sigma^2}{\gamma+\kappa}\right)^b \left(\frac{\kappa-\gamma}{\kappa+\gamma}\right)^n, 
\end{equation*}
where $L_n^{(\alpha)}(x)$ are the generalized Laguerre polynomials and $b$ and $\gamma$ as defined in \eqref{gamb}.

The CIR eigenfunctions are continuous and have a bound independent of $n$ on each compact interval $K=[a,b] \subset (l,r)$, i.e. $|\varphi_n(x) |\leq C_K$ for all $n$, where the constant $C_K$ is independent of $n$, since by inequality (27a) on p.53 of \citet{niki}, for any compact interval in $I$, the CIR eigenfunctions satisfy the bound $\left| \varphi_n(x) \right| \leq C n^{-1/4}, $ where the constant $C$ is independent of $n$ (but depends on the interval).

The quantities \eqref{pieqn} and \eqref{pneqn} in the CIR model can be calculated as follows: 
\begin{equation} \label{cirpi}
\pi_{m,n}(x,y)=
\left( \frac{\sigma^2}{2\gamma} \right)^{b-1} \frac{N_n N_m}{\gamma} \left[ a_{n,m}^{(b-1)}\left( \frac{2\gamma y}{\sigma^2} \right)- a_{n,m}^{(b-1)}\left( \frac{2\gamma x}{\sigma^2} \right) \right],
\end{equation}
\begin{equation} \label{cirp}
p_n(x,y)=A(\delta)\frac{N_n}{\gamma}\left( \frac{\sigma^2}{2\gamma}\right)^{b-1} \left( b_n^{(b-1)}\left(s,\frac{2\gamma y}{\sigma^2}\right)-b_n^{(b-1)}\left(s,\frac{2\gamma x}{\sigma^2}\right)\right),
\end{equation}
where we introduced the following notation:
$$s=\frac{B(\delta)\sigma^2}{2 \gamma}+\frac{\kappa+\gamma}{2\gamma},$$
$$a_{n,m}^{(\alpha)}(x) = \int_0^x L_n^{(\alpha)}(y) L_m^{(\alpha)}(y) e^{-y} y^{\alpha}\,  \mathrm{d}y,\quad
b_n^{(\alpha)}(s,x)=\int_0^x y^{\alpha} e^{-sy} L_n^{(\alpha)}\left( y \right) \, \mathrm{d}y.$$
In the calculation of \eqref{cirp} we used the explicit expression for the CIR zero-coupon bond pricing function \eqref{cirzerobond} as in Eq.\eqref{pnzerobond}, rather than its eigenfunction expansion.

The quantities $a_{n,m}^{(\alpha)}(x)$ and $b_n^{(\alpha)}(s,x)$ with $\alpha=b-1>-1$ can be efficiently computed via the following recursion.
\begin{propo}
\label{cirpropoab}
Suppose that $\alpha > -1$.  The coefficients $a_{n,m}^{(\alpha)}(x)$ are computed as follows for all $x>0$.
For $n\geq 1$, $m\geq 1$, $m\neq n$,
\begin{equation} 
a_{m,n}^{(\alpha)}(x) = \frac{e^{-x} x^{\alpha+1}}{m-n}\left( L_{n}^{(\alpha)}(x)L_{m-1}^{(\alpha+1)}(x)-L_{m}^{(\alpha)}(x)L_{n-1}^{(\alpha+1)}(x) \right).
\label{cira1}
 \end{equation}
For $n\geq 1$,
\begin{equation}
a_{0,n}^{(\alpha)}(x)=\frac{1}{n} e^{-x} x^{\alpha+1} L_{n-1}^{(\alpha+1)}(x).
\label{cira2}
\end{equation}
For $m=n$,
\begin{equation}
a_{0,0}^{(\alpha)} = \gamma(\alpha+1,x), \quad a_{n,n}^{(\alpha)}(x)=\frac{1}{n} \left[ L_n^{(\alpha)}(x) L_{n-1}^{(\alpha+1)}(x)e^{-x} x^{\alpha+1}+a_{n-1,n-1}^{(\alpha+1)}(x) \right],\quad n\geq 1,
\label{cira3} \end{equation}
where $\gamma (\alpha+1,x)=\int_0^x e^{-y}y^\alpha \, \mathrm{d}y$ is the lower incomplete gamma function.

The coefficients $b_n^{(\alpha)}(s,x)$ are computed recursively as follows for all $x>0$.
\begin{equation} 
b_0^{(\alpha)}(s,x)= \frac{1}{s^{\alpha+1}} \gamma (\alpha+1,sx),\quad  b_n^{(\alpha)}(s,x)= \frac{1}{n}e^{-sx}x^{\alpha+1} L_{n-1}^{(\alpha+1)}(x) + \frac{s-1}{n} b_{n-1}^{(\alpha+1)}(x),\quad n\geq 1, \label{cirb1propo}
\end{equation}
\end{propo}
\begin{proof}
See the Appendix.
\end{proof}

The Laguerre polynomials of degree $\alpha$  satisfy the following classical recursion \citep[Eq. 4.18.1]{lebedev}
$$
L_n^{(\alpha)}(x) = \left(2+ \frac{\alpha-1-x}{n}\right) L_{n-1}^{(\alpha)}(x) - \left(1+\frac{\alpha-1}{n} \right) L_{n-2}^{(\alpha)}(x),\quad n\geq 2,
$$ 
and $L_0^{(\alpha)}(x) = 1$, $L_1^{(\alpha)}(x)=-x+\alpha+1$.  Then for any $N$ the quantities $\{a_{n,n}^{(\alpha)}(x) , 0\leq n \leq N \}$ can be efficiently computed recursively in the following order:
\begin{itemize}
\item $a_{0,0}^{(\alpha+N)}(x)$
\item $a_{0,0}^{(\alpha+N-1)}(x)$, $a_{1,1}^{(\alpha+N-1)}(x)$
\item $\vdots$
\item $a_{0,0}^{(\alpha+1)}(x)$, $a_{1,1}^{(\alpha+1)}(x)$, $\cdots$, $a_{N-1,N-1}^{(\alpha+1)}(x)$
\item $a_{0,0}^{(\alpha)}(x)$, $a_{1,1}^{(\alpha)}(x)$, $\cdots$, $a_{N-1,N-1}^{(\alpha)}(x)$, $a_{N,N}^{(\alpha)}(x)$
\end{itemize}
The computation of $a_{n,m}^{(\alpha)}(x)$ with $n\not=m$ can be done directly using \eqref{cira1} and \eqref{cira2} and the recursion for the Laguerre polynomials.
The quantities $\{b_n^{(\alpha)}(x), 0\leq n \leq N \}$ can be efficiently computed recursively in the order:
\begin{itemize}
\item $b_0^{(\alpha+N)}(x)$
\item $b_0^{(\alpha+N-1)}(x)$, $b_1^{(\alpha+N-1)}(x)$
\item $\vdots$
\item $b_0^{(\alpha+1)}(x)$, $b_1^{(\alpha+1)}(x)$, $\cdots$, $b_{N-1}^{(\alpha+1)}(x)$
\item $b_0^{(\alpha)}(x)$, $b_1^{(\alpha)}(x)$, $\cdots$, $b_{N-1}^{(\alpha)}(x)$, $b_N^{(\alpha)}(x)$
\end{itemize}

For the SubCIR model the explicit bond pricing formula similar to \eqref{cirzerobond} is not available, and we use the eigenfunction expansion \eqref{subzerobond} of the SubCIR zero-coupon bond price as in Eq.\eqref{generalzerobond}. The expression \eqref{cirp} is then replaced with:
\begin{equation}
p_n(x,y)=\sum_{m=0}^\infty
p_m e^{-\phi(\lambda_m )\delta}  \left( \frac{\sigma^2}{2\gamma} \right)^{b-1} \frac{N_nN_m}{\gamma} \left[a_{n,m}^{(b-1)}  \left( \frac{2\gamma y}{\sigma^2} \right) - a_{n,m}^{(b-1)}\left( \frac{2\gamma x}{\sigma^2} \right) \right].
\end{equation} 
In Theorem 3.1 $\lambda_n$ are now the eigenvalues of the SubCIR model related by $\phi(\lambda_n)$ to the eigenvalues of the CIR diffusion model.

\begin{remark} In the limiting case $x=\infty$ we have 
$$
a_{m,n}^{(\alpha)}(\infty)=\frac{\gamma}{N_nN_m} \left( \frac{2\gamma}{\sigma^2} \right)^{\alpha}  \delta_{m,n},\quad b_n^{(\alpha)}(s,\infty)=\frac{\Gamma(\alpha+n+1) (s-1)^n}{n! s^{\alpha+n+1}} .
$$
due to the orthogonality of Laguerre polynomials and the integral identity \citep[p.809]{grad} 
$$
\int_0^{\infty} e^{-sy} y^\alpha L_n^{(\alpha)} (y) \mathrm{d}y =  \frac{\Gamma(\alpha+n+1)(s-1)^n}{n! s^{\alpha+n+1}},
$$ 
where $\alpha>-1$, $s>0$, $n=0,1,2,....$ Using these coefficients, the recursion for the expansion coefficients in Theorem 3.1 simplifies in the case of callable bonds with no put option (in that case $x_i=\infty$ and there is no term with $K_i^p$ in Eq.(20)). 
\end{remark}
 
 \subsection{Vasicek and SubVasicek}
\label{sec:vasicekmodel}
 
In Vasicek model \citep{vasicek}, the short rate follows the OU diffusion with drift $\mu(x)=\kappa(\theta-x)$ with $\kappa>0$ and $\theta>0$ and constant volatility $\sigma>0$.
In this case $I={\mathbb R}$ and $r(x)=x$, both boundaries at plus and minus infinity are unattainable natural boundaries, and the process can get negative. However, when $\theta$ and the initial state $x_0$ are sufficiently above zero and $\kappa>0$ is sufficiently large, the probability of the rate falling below zero is relatively small due to mean reversion pulling the process back towards the positive long run mean as it approaches zero from above. 
The Vasicek speed density is a Gaussian density $m(x)=\frac{2}{\sigma^2}e^{-\frac{\kappa(\theta-x)^2}{\sigma^2}}$.

The celebrated Vasicek zero-coupon bond pricing formula has the same exponential affine form as the CIR \eqref{cirzerobond} with
$$ B(t)=\frac{1}{\kappa}\left( 1- e^{-\kappa t} \right), \quad
A(t)=\exp\left\{\frac{1}{\kappa^2} (B(t)-t)(\kappa^2 \theta-\sigma^2/2) -\frac{\sigma^2 B(t)^2}{4\kappa} \right\}.$$

The eigenfunction expansion \eqref{bondprice} of the Vasicek zero-coupon bond pricing function is given in
\cite{gorovoi}. In this case the eigenfunctions, eigenvalues, and the expansion coefficients for the unit payoff are:
$$\lambda_n = \theta-\frac{\sigma^2}{2\kappa^2}+\kappa n,$$
$$\varphi_n(x) = N_n e^{-a \xi - \frac{a^2}{2}}H_n(\xi+a),\quad \xi:=\frac{\sqrt{\kappa}}{\sigma}(x-\theta), \quad a:=\frac{\sigma}{\kappa^{3/2}},\quad
N_n= \sqrt{\sqrt{\frac{\kappa}{\pi}} \frac{\sigma}{2^{n+1} n !} },$$
$$p_n = \frac{2}{\sigma}\sqrt{\frac{\pi}{\kappa}}N_n a^n e^{-\frac{a^2}{4}},$$
where $H_n(x)$ are Hermite polynomials.

The Vasicek eigenfunctions are continuous and have a bound independent of $n$ on each compact interval $K=[a,b] \subset (l,r)$, i.e. $|\varphi_n(x) |\leq C_K$ for all $n$, where the constant $C_K$ is independent of $n$, since by inequality (28a) on p.53 of \citet{niki}, for any compact interval in $I$, the Vasicek eigenfunctions satisfy the bound $\left| \varphi_n(x) \right| \leq C n^{-1/4}, $ where the constant $C$ is independent of $n$.

 The quantities \eqref{pieqn} and \eqref{pneqn} in the Vasicek model can be calculated as follows: 
\begin{equation} \label{vasicekpi}
\pi_{m,n}(x,y)= \frac{2N_n N_m}{\sigma \sqrt{\kappa}} \left[a_{n,m}\left(\frac{\sqrt{\kappa}}{\sigma}(y-\theta)+a \right)- a_{n,m}\left(\frac{\sqrt{\kappa}}{\sigma}(x-\theta)+a \right) \right],
\end{equation}
\begin{equation} \label{vasicekp}
p_n(x,y)=\frac{2A(\delta)N_n}{\sigma \sqrt{\kappa}} e^{-\frac{a^2}{2}-B(\delta)(\theta-\frac{a\sigma}{\sqrt{\kappa}})} \left[b_n\left(s,\frac{\sqrt{\kappa}}{\sigma}(y-\theta)+a\right)-b_n\left(s,\frac{\sqrt{\kappa}}{\sigma}(x-\theta)+a\right)\right],
\end{equation}
where we introduced the following notation:
$$s:=-\frac{B(\delta) \sigma}{\sqrt{\kappa}}+a,$$
$$a_{n,m}(x): = \int_{-\infty}^x e^{-y^2}H_n(y) H_m(y) \,\mathrm{d}y  ,\quad  b_n(s,x):=\int_{-\infty}^x e^{sy-y^2} H_n(y) \, \mathrm{d}y.$$
In the calculation of \eqref{vasicekp} we used the explicit expression for the Vasicek zero-coupon bond pricing function as in \eqref{pnzerobond}, rather than its eigenfunction expansion.

The quantities $a_{n,m}$ and $b_n$ can be computed efficiently.  The coefficients of the Hermite polynomial $H_n(x)$ can be computed from the recursive equation \citep[Eq. 4.10.1]{lebedev} $$H_n(x) = 2xH_{n-1}(x) - 2(n-1) H_{n-2} (x),\quad n\geq 2,\quad  H_0(x)=1,\quad H_1(x)=2x. $$ 
\begin{propo}
\label{vasicekpropoab}
The quantities $a_{m,n}(x)$ can be computed as follows.
For $m\neq n$
 \begin{equation} a_{n,m}(x) = \frac{H_n(x)H_{m+1}(x)-H_m(x)H_{n+1}(x)}{2(m-n)}e^{-x^2}. \label{propovasiceka1} \end{equation}
$a_{n,n}(x)$ can be computed recursively as follows:
 \begin{equation} 
a_{0,0}(x)=\sqrt{\pi}\Phi(\sqrt{2}x),\quad a_{n,n}(x) = -H_{n-1}(x)H_n(x) e^{-x^2} +2na_{n-1,n-1}(x),\quad n\geq 1,  \label{propovasiceka2} \end{equation}
where $\Phi(x)$ is the standard normal cumulative distribution function.

The quantities $b_n(s,x)$ can be computed as follows.
\begin{equation} 
b_0(s,x)=\frac{1}{2} e^{\frac{s^2}{4}} \sqrt{\pi}\left(\mathrm{Erf}\left(\frac{1}{2}(2x-s)\right)+1\right), \label{propovasicekb2}
\end{equation}
where $\mathrm{Erf}(x)$ is the error function, and
\begin{equation}
b_n(s,x) = -e^{sx-x^2} H_{n-1} (x) +sb_{n-1}(s,x),\quad n\geq 1. \label{propovasicekb1}
\end{equation}
\end{propo}
\begin{proof}
See the Appendix.
\end{proof}

For the SubVasicek model the explicit bond pricing formula similar to \eqref{zerobondexp} is not available, and we use the eigenfunction expansion \eqref{subzerobond} of the SubVasicek zero-coupon bond price instead. The expression \eqref{vasicekp} is then replaced with:
\begin{equation}
p_n(x,y)=\sum_{m=0}^\infty p_m e^{-\phi(\lambda_m) \delta} \frac{2N_n N_m}{\sigma \sqrt{\kappa}} \left[a_{n,m}\left(\frac{\sqrt{\kappa}}{\sigma}(y-\theta)+a \right) -a_{n,m}\left(\frac{\sqrt{\kappa}}{\sigma}(x-\theta)+a \right)\right].
\end{equation} 
In the recursion \eqref{coeffrecurthm} $\lambda_n$ are the eigenvalues of the SubVasicek model related by $\phi(\lambda_n)$ to the eigenvalues of the Vasicek diffusion model.

\begin{remark} In the limiting case $x=\infty$ we have 
$$
a_{m,n}(\infty)=\frac{\sigma \sqrt{\kappa}}{2N_n N_m} \delta_{m,n},\quad b_n(s,\infty)= e^{s^2/4}\sqrt{\pi} (-s)^n
$$
due to the orthogonality of Hermite polynomials  and the integral identity \citep[Eq. 7.374.6, p.803]{grad} $$\int_{-\infty}^{\infty} e^{-(x+y)^2} H_n(x) \,\mathrm{d}x = \sqrt{\pi} (-2y)^n.$$ These coefficients can be used to evaluate callable bonds without the put option, similar to remark 4.1 for the CIR.
\end{remark}

\subsection{The 3/2 and Sub-3/2 Model}

In this model, the short rate process is a diffusion on $(0,\infty)$ with infinitesimal parameters $\sigma(x) = \sigma x^{3/2},$ $\mu(x) = \kappa (\theta-x)x,$  $r(x)=x ,$
where $\kappa$, $\theta$, and $\sigma$ are positive constant parameters.  This process was proposed by \cite{cir} as a model for the inflation rate. \cite{ahngao} propose this process as a model for the short rate and show that this model is empirically more plausible than the square-root model.

Let $\alpha=\frac{\kappa}{\sigma^2}+1$, $\beta=\frac{2\kappa\theta}{\sigma^2}$, $m=\sqrt{\left(\frac{\kappa}{\sigma^2}+\frac{1}{2} \right)^2+ \frac{2}{\sigma^2}}.$
The speed density for this model is $m(x) = \frac{2}{\sigma^2}x^{-2\alpha-1}e^{-\frac{\beta}{x}}.$
The eigenfunction expansion \eqref{bondprice} of the zero-coupon bond pricing bond is given in \cite{linet1}.  In this case the eigenfunctions, eigenvalues, and the expansion coefficients for the unit payoff are:
$$\lambda_n = \kappa \theta(n+m-\alpha+1/2),$$
$$\varphi_n(x) = N_n x^{\alpha-m-1/2} L_n^{(2m)}\left(\frac{\beta}{x} \right), \quad N_n= \sqrt{\frac{\sigma^2 \beta^{2m+1} n!}{2\Gamma(2m+n+1)} },$$ 
$$p_n = \frac{2}{\sigma^2}N_n \beta^{-\alpha-m-\frac{1}{2}} \frac{\Gamma(\alpha+m+1/2) \Gamma(m+n-\alpha+1/2)}{n! \Gamma(m-\alpha+1/2)} .$$
 
The 3/2 eigenfunctions are continuous on $I$ and have a bound independent of $n$ on each compact interval $K=[a,b] \subset (l,r)$, i.e. $|\varphi_n(x) |\leq C_K$ for all $n$, where the constant $C_K$ is independent of $n$, since by inequality (27a) on p.53 of \citet{niki}, for any compact interval in $I$, the 3/2 eigenfunctions satisfy the bound $\left| \varphi_n(x) \right| \leq C n^{-1/4}, $ where the constant $C$ is independent of $n$.

The quantities \eqref{pieqn} and \eqref{pneqn} in the 3/2 and Sub-3/2 model can be calculated as follows: 
\begin{equation} \label{threetwopi}
\pi_{k,n}(x,y)=  \frac{2N_n N_k}{\sigma^2 \beta^{2m+1}} \left[ a_{n,k}^{(2m)} \left(\frac{\beta}{x} \right)- a_{n,k}^{(2m)} \left(\frac{\beta}{y} \right) \right],
\end{equation}
\begin{equation} \label{threetwop}
p_n(x,y)=\sum_{k=0}^\infty p_k e^{-\lambda_k \delta} \frac{2N_n N_k}{\sigma^2 \beta^{2m+1}} \left[ a_{n,k}^{(2m)} \left(\frac{\beta}{x} \right)- a_{n,k}^{(2m)} \left(\frac{\beta}{y} \right) \right],
\end{equation}
where we introduced the following notation:
$$a_{n,k}^{(\alpha)}(x) = \int_0^x L_n^{(\alpha)}(y) L_k^{(\alpha)}(y) e^{-y} y^{\alpha} \mathrm{d}y.$$
For the Sub-3/2 model,  $\lambda_n$ are the eigenvalues of the Sub-3/2 model related by $\phi(\lambda_n)$ to the eigenvalues of the 3/2 diffusion model.

The expression for $a_{n,k}(x)$ is same as the one for the CIR model.  Hence, the quantities $a_{n,k}(x)$ can be computed using the method given in the CIR section.

 \section{Computational Results}
\label{sec:result}

This section shows our computational results for CIR, Vasicek, SubCIR, and SubVasicek models.  We consider the callable bond example that has been extensively used in the literature starting from \cite{bw} and including \cite{dfvl}, \cite{bbkl}, and \cite{frutos}, as the test case to compare computational performance of a number of computational approaches to the callable bond valuation. The callable bond was issued by Swiss Confederation in 1987 with maturity in 2012.  At the time of valuation considered in \cite{bw} and the subsequent papers, the remaining time to maturity of the bond was $t_k=20.172$ years with $k=21$ remaining annual coupons of $4.25\%$ per annum.  The notice period is 2 months, $\delta=0.1666$.  The protection period is $t_{k^*}=10.172$ with $k^*=11$.  There are ten early exercise dates, $t_{11}=10.172$, $t_{12}=11.172$,..., $t_{20}=19.172$.  The call prices corresponding to these dates are given in Table \ref{callprice}. The bond did not include a put option.  We use the values of the parameters $\kappa$, $\sigma$, $\theta$  for the CIR and Vasicek models estimated in \cite{bw} and used in the subsequent papers in the literature. They are given in Table \ref{parameters}.   For the SubCIR and SubVasicek models, we used the same parameter values for the underlying CIR and Vasicek diffusions, while specifying the subordinator to be the inverse Gaussian (IG) subordinator with drift (IG L\'{e}vy measure given in \eqref{nugauss}).  The IG parameter $\nu$ was set to 1.  The subordinator drift $\gamma$ and the IG parameter $\mu$ were chosen so that ${\mathbb E}[T_t]=t$ to normalize the time change.  For the jump-diffusion case, we used $\gamma=0.5$ and $\mu=0.5$.  For the pure jump case, we used $\gamma=0$ and $\mu=1$.


In the process of finding the break-even point $x_i^c$ at each step of the recursion, the infinite series for the continuation value given in \eqref{vholdeig1} needs to be truncated at some finite level.  At time $t_0$, the series in \eqref{v0thm} also needs to be truncated.  In the recursion formula for the expansion coefficients in \eqref{coeffrecurthm}, only the previous expansion coefficients that were calculated at the previous step are used in the eigenfunction expansion.  For the subordinated models, we also compute the zero-coupon bonds by the eigenfunction expansions that are also truncated at some finite level. 
We used an adaptive truncation strategy that truncated the expansion after a user-specified relative error tolerance $\epsilon$ was reached in each instance of the series evaluation by comparing with $\epsilon$ the ratio of the next term and the sum of the next two terms relative to the sum of all the previous terms.
 

In order to find the break-even point at each decision point in time, the bisection method was used. 
The break-even short rates are shown in table \ref{breakeven1} (JD stands for jump-diffusion, and PJ for pure jump).  For Vasicek and CIR models, $r(x)=x$, so the short rate is equal to the state value. For subordinated models, the short rate is given by the function $r^\phi(x)$ of the state variable.  For the CIR model, we start by checking the boundary at zero to see if the expected discounted value of the strike is greater than the continuation value, in which case there is no non-negative break-even point at that decision time instance.  Otherwise, there is a unique non-negative break-even point for the CIR model.  For the Vasicek model, there always is a unique break-even point.  The break-even point was found by the bisection method until the length of the search interval became less than $10^{-7}$.

Table \ref{conv1}  shows computational results for all the models considered in this section with the initial short rate $r=0.05$.  The first column indicates the absolute pricing error in pricing the callable bond.   During the process of finding the break-even point, the truncation level is determined for each evaluation of the continuation value.  The second column gives the average truncation level $N$ at each decision point at time $\tau_{20}$, $\tau_{19}$, ..., $\tau_{11}$, and at $t_0$ (the average of truncations levels as determined by our adaptive truncation algorithm in evaluating the expansion of the continuation value needed for each step of the bisection algorithm).  The third column shows the maximum truncation level $N$ at those times.  The fourth column shows the CPU time of our algorithm implemented in C using the GNU Scientific Library (GSL) and compiled with gcc and executed on a 2.4 GHz Intel Core i3 370M processor.  The CPU time includes the time taken for any precomputations required.  For CIR and Vasicek models, the CPU time to price the callable bond to approximately five correct decimal points (convergence of $10^{-5}$ in the tables) was about one millisecond.  For comparison, \cite{frutos} reported CPU times of 0.75 seconds using Matlab on a 3GHz processor, while \cite{bbkl} reported CPU times of 2 to 3 seconds using C on a 2.0 GHz Pentium 4 processor. Thus, the eigenfunction expansion approach is approximately three orders of magnitude faster in this instance of pricing the callable bond.

For the subordinated models with jumps, jump-diffusion models required slightly longer CPU times than pure diffusion models, while pure jump models required slightly longer times than jump-diffusions. This is due to the replacement of the diffusion eigenvalues $\lambda_n$ with eigenvalues $\phi(\lambda_n)$ of subordinated processes that slows down the eigenvalue growth and, hence, requires more terms in the expansions, as evidenced in Table 4. Still, the algorithm reached the pricing error of under $10^{-5}$ in 2.2 and 2.5 milliseconds under pure jump CIR and Vasicek models. 
Tables \ref{cirsummary} and \ref{vasiceksummary} show the computed values of the callable bond in comparison to other methods.  The columns BW, DFVL, BBKL, and F refer to \cite{bw}, \cite{dfvl}, \cite{bbkl}, and \cite{frutos}.  Table  \ref{subcirsummary1} shows the results for the subordinated models in the present paper.
{\em We stress that neither of the alternative approaches in the literature is capable of handling jump-diffusion and pure jump models with state dependent jumps. The remarkable advantage of the eigenfunction expansion method is that it is entirely straightforward to move from pure diffusion models to jump-diffusion and pure jump models obtained by subordination by simply replacing the diffusion eigenvalues $\lambda_n$ with subordinate eigenvalues $\phi(\lambda_n)$.}

While Tables 1-7 provide results for the bond with the call option only to facilitate comparisons with the literature, Tables 8-10 provide the corresponding results for the bond that is both callable and putable. To generate this example of a bond with both options, we added the put option to the callable bond considered previously in this section. Table \ref{putprice} gives the put prices we have assumed. Table \ref{breakevencallput} presents results for break-even short rates for call and put options under the range of short rate models considered in this paper. Table \ref{summarycallput} presents the corresponding prices of the bond with both call and put options.

\section{Conclusion}
\label{sec:conclusion}

This paper proposed an efficient method to evaluate bonds with embedded options under a wide class of interest rate models, including the popular short rate diffusion models, as well as their time changed versions with jumps. The method is based on the eigenfunction expansion of the pricing operator. Given the set of call and put dates, the callable and putable bond pricing function is the value function of a stochastic game with stopping times. Under some technical conditions, it is shown to have an eigenfunction expansion in eigenfunctions of the pricing operator with the expansion coefficients determined through a backward recursion. For CIR and Vasicek the method is orders of magnitude faster than the alternative approaches in the literature. In contrast to the alternative approaches in the literature that have so far been limited to diffusions, the method is equally applicable to short rate jump-diffusion and pure jump models constructed from diffusion models by Bochner's subordination with a L\'{e}vy subordinator. In future work we plan to apply the eigenfunction expansion method of this paper to convertible bonds, where the stock price process is the stochastic variable driving the conversion and call decisions.

\section{Appendix}
\label{sec:appendix}

\begin{proof}[Proof of Theorem \ref{section3theorem}] 
(i) For any $f\in L^2(I,m)$,  $\mathcal{P}_tf \in L^2(I,m)$ for any $t\geq 0$.  Since $m$ is a finite measure on $I$, $1 \in L^2(I,m)$, so $P(t,x)\in L^2(I,m)$ for any $t\geq 0$.  Then by \eqref{vk} to \eqref{vtau}, $C^{k-1}(x)$ and $V^{k-1}(x)$ are in $L^2(I,m)$.  By induction using \eqref{continuation}, \eqref{vtau}, and \eqref{v0}, it can be shown that the value function $V^0(x)$ and the value functions $V^i(x)$ and the continuation value function $C^i(x)$ are in $L^2(I,m)$ for all $i=k^*, ..., k-1$.

(ii) The expressions for the continuation value function and the value function at the time of the bond issue are given in \eqref{continuation} and \eqref{v0}.  By part (i), the eigenfunction expansion can be obtained from \eqref{eigenexpansion} or \eqref{subeigenexpansion}, where $c^{i+1}_n = (V^{i+1},\varphi_n)$, $i=k^*, ..., k-1$, and $c^{k^*}_n = (V^{k^*},\varphi_n)$ for $n=0,1,2,...$.

(iii) Under the conditions given in section 2.1 or 2.2, the eigenfunction expansion for the density given in \eqref{transitioneig1} or \eqref{transitioneig2} holds.  Then for $f \in L^2(I,m)$ and any $x\in I$,

\begin{eqnarray*}
\mathcal{P}_t f(x) &=& \int_I f(y)p_t(x,y)m(y) \, \mathrm{d}y = \int_I f(y) \sum_{n=0}^\infty e^{-\lambda_n t} \varphi_n(x) \varphi_n(y) m(y) \, \mathrm{d}y \\
&=& \sum_{n=0}^\infty e^{-\lambda_n t }\varphi_n(x) \int_I f(y) \varphi_n(y) m(y)\, \mathrm{d}y = \sum_{n=0}^\infty f_n e^{-\lambda_n t} \varphi_n(x),
\end{eqnarray*}
where $f_n=(f,\varphi_n).$
The interchange in the third equality is justified by the Dominated Convergence Theorem with the dominant function $\sum_{n=0}^\infty e^{-\lambda_n t} \left| \varphi_n(x) f(y) \varphi_n(y) m(y) \right|$:
\begin{eqnarray*}
\sum_{n=0}^\infty \int_I e^{-\lambda_n t} \left| \varphi_n(x) f(y) \varphi_n(y) m(y) \right| \, \mathrm{d}y &\leq &  \sum_{n=0}^\infty e^{-\lambda_n t } \left|\varphi_n(x) \right| \|f\|_{L^2} \|\varphi_n\|_{L^2} \\
 &=& \| f\|_{L^2} \sum_{n=0}^\infty e^{-\lambda_n t}\left| \varphi_n(x) \right| <\infty
\end{eqnarray*}
The first inequality follows from the Cauchy-Schwartz inequality, and the last inequality follows from the bounds on eigenfunctions described in section 2.1 or 2.2 (e.g. $\left| \varphi_n(x) \right| \leq e^{\lambda_n t/2} \sqrt{p_t(x,x)}$ or $\left| \varphi_n(x) \right| \leq C_K$) and the trace class condition \eqref{traceclass} or \eqref{subuniform}.
Hence, the eigenfunction expansion converges pointwise to $\mathcal{P}_t f(x)$ for any $x\in I$.  

The eigenfunction expansions converge uniformly in $x$ on compacts in $I$ by the following:
  Let $K$ be any compact subset of $I$.  For $x\in K$,
\begin{equation*}
\left| \sum_{n=M}^\infty f_n e^{-\lambda_n t}\varphi_n(x)   \right|  \leq  \sum_{n=M}^\infty \left| f_n e^{-\lambda_n t} \varphi_n(x) \right|  \leq  \|f \|_{L^2} \sum_{n=M}^\infty e^{-\lambda_n t} \left| \varphi_n(x) \right|.
\end{equation*}
The last expression goes to 0 as $M$ goes to $\infty$ by the bounds on $\left| \varphi_n(x) \right|$ and the trace class condition.  Therefore, the eigenfunction expansions converge uniformly in $x$ on compacts in $I$.

(iv) By \eqref{vk}, the values of $c^k_n$ are given by the coefficients of the eigenfunction expansion of the zero-coupon bond:
$$c^k_n=(1+C)p_n, \quad n=0,1,2,...$$
Suppose that we know the values of the coefficients $c_n^{i+1}$, $n=0,1,2,...$.
Denote $1_{(x,y)}$ as an indicator function that is 1 on the interval $(x,y)$ and 0 otherwise.

\begin{eqnarray}
c_n^i &=& \left(V^i,\varphi_n\right) + Cp_n e^{-\lambda_n \delta} \nonumber \\
  &=& \left(K_i^c P(\delta) 1_{(l,x^c_i)} ,\varphi_n\right)+ \left( \sum_{m=0}^\infty c_m^{i+1} e^{-\lambda_m h_i} \varphi_m 1_{(x_i^c, x_i^p)},\varphi_n\right) 
+\left(K_i^p P(\delta) 1_{(x_i^p,r)},\varphi_n\right) + Cp_n e^{-\lambda_n \delta}\nonumber\\
&=& K^c_ip_n(l,x^c_i) +\left( \sum_{m=0}^\infty c_m^{i+1} e^{-\lambda_m h_i} \varphi_m ,\varphi_n 1_{(x_i^c, x_i^p)}\right) 
+K^p_i p_n(x^p_i,r) + Cp_n e^{-\lambda_n \delta}  \nonumber\\
&=& K^c_ip_n(l,x^c_i)+\left( \sum_{m=0}^\infty c_m^{i+1} e^{-\lambda_m h_i} \varphi_m ,\sum_{m=0}^\infty \pi_{m,n}(x_i^c,x_i^p) \varphi_m \right) 
+K^p_i p_n(x^p_i,r) + Cp_n e^{-\lambda_n \delta}  \nonumber\\
&=& K^c_ip_n(l,x^c_i) + \sum_{m=0}^\infty c_m^{i+1}e^{-\lambda_m h_i} \pi_{m,n}(x_i^c,x_i^p)  
   + K^p_i p_n(x^p_i,r) + Cp_n e^{-\lambda_n \delta} \nonumber.
\end{eqnarray}
The last equality follows from $(f,g) = \sum_{n=0}^\infty f_n g_n,$ where $f_n=(f,\varphi_n)$ and $g_n=(g,\varphi_n)$ for $f,g \in L^2(I,m)$.
\end{proof}

\begin{proof}[Proof of Proposition \ref{atmost}]  
Suppose that \eqref{comparison} holds.
From this we get that, with probability one, $$e^{-\int_0^t r(X_1(s)) \, \mathrm{d}s} > e^{-\int_0^t r(X_2(s)) \, \mathrm{d}s}.$$  Hence, the zero-coupon bond function $P(t,x)$ is a positive decreasing function.

We can also show that the continuation value function $C^i(x)$ and bond value function $V^i(x)$  are positive decreasing functions of $x$ for all $i=k^*, ..., k-1.$  For $i=k-1$, $C^{k-1}(x)$ is a constant multiple of a zero-coupon bond function by \eqref{vk} and \eqref{continuation}, so $C^{k-1}(x)$ is a positive decreasing function.  By \eqref{vtau}, the bond value function $V^i(x)$ is a positive decreasing function if $C^i(x)$ is a positive decreasing function since we already showed that zero-coupon bond functions are positive decreasing functions.  It remains to show that for $i=k^*, ..., k-2$ the continuation function $C^i(x)$ is a positive decreasing function given that $V^{i+1}(x)$ is a positive decreasing function.  This is shown by the following.  By \eqref{comparison}, $$e^{-\int_{0}^{t} r(X_1(s)) \, \mathrm{d}s} V^{i+1}(X_1(t))>e^{-\int_{0}^{t} r(X_2(s)) \, \mathrm{d}s} V^{i+1}(X_2(t))$$ for $X_1(0) < X_2(0).$

We show that there can be at most one solution $x$ such that $$K P(\delta,x) = C^i(x),$$ where $K$ is a constant (either $K_i^c$ or $K_i^p$), for each $i=k^*,...,k-1$. 
 It is first shown that there is at most one solution to the equation given by $$KP(\delta,x) = C^{k-1}(x).$$  Suppose that $x_{k-1}$ is a solution to the equation.
By \eqref{continuation}, 
\begin{eqnarray*} 
C^{k-1}(x) &=& (1+C)E\left[e^{-\int_{\tau_{k-1}}^{t_k} r(X(s)) \, \mathrm{ds}} \middle\vert  X(\tau_{k-1})=x\right]\\
&=& (1+C)E\left[e^{-\int_{\tau_{k-1}}^{t_{k-1}} r(X(s)) \, \mathrm{ds}} g(X(t_{k-1})) \middle\vert X(\tau_{k-1})=x\right]\\
&=& (1+C) P(t_{k-1}-\tau_{k-1},x) E^{t_{k-1}}\left[g(X(t_{k-1})) \middle\vert X(\tau_{k-1})=x\right],
\end{eqnarray*}  
where $g(x)=E\left[e^{-\int_{t_{k-1}}^{t_k} r(X(s)) \, \mathrm{ds}} \middle\vert X(t_{k-1})=x\right]$ is a decreasing function and $E^{t_{k-1}}$ denotes the expectation under the $t_{k-1}$-forward adjusted measure.  Denote the probability measure under the $t_{k-1}$-forward adjusted measure as $\mathbb{Q}_{k-1}$.  $\mathbb{Q}_{k-1}$ and $\mathbb{P}$ are equivalent probability measures, so $\mathbb{P}[X_1(t) \leq X_2(t) \textrm{ for all } t\geq 0] =1$ implies that $\mathbb{Q}_{k-1}[X_1(t) \leq X_2(t) \textrm{ for all } t\geq 0] =1$.  Then  $E^{t_{k-1}}\left[g(X(t_{k-1})) \middle\vert X(\tau_{k-1})=x\right]$ is a decreasing function of $x$.  At $x=x_{k-1}$, 
$$KP(t_{k-1}-\tau_{k-1},x)=(1+C)P(t_{k-1}-\tau_{k-1},x)E^{t_{k-1}}\left[g(X(t_{k-1})) \middle\vert X(\tau_{k-1})=x\right].$$
Then $K=(1+C)E^{t_{k-1}}\left[g(X(t_{k-1})) \middle\vert X(\tau_{k-1})=x_{k-1}\right]$, so $$K>(1+C)E^{t_{k-1}}\left[g(X(t_{k-1})) \middle\vert X(\tau_{k-1})=x\right]$$ for all $x>x_{k-1}$.  Then $K P(t_{k-1}-\tau_{k-1},x) > C^{k-1}(x)$ for all $x>x_{k-1}$.

It is shown next that there is at most one solution to the equation given by $$K P(\delta,x) = C^i(x),$$ for each $i=k^*,...,k-2$, where $K$ is a constant (either $K_i^c$ or $K_i^p$).  Let $x_i$ be such that $K P(\delta,x_i) = C^i(x_i)$.  
By \eqref{continuation},
\begin{eqnarray*}
 C^i(x) &=& E\left[e^{-\int_{\tau_i}^{\tau_{i+1}} r(X(s)) \, \mathrm{d}s} V^{i+1}(X(\tau_{i+1})) \middle\vert X(\tau_i)=x\right]\\
 &=& E\left[e^{-\int_{\tau_i}^{t_i} r(X(s)) \, \mathrm{d}s} g(X(t_i)) \middle\vert X(\tau_i)=x\right] \\
 &=& P(\delta,x)E^{t_i}\left[g(X(t_i))\middle\vert X(\tau_i)=x\right] , 
\end{eqnarray*}
where $g(x)=E\left[e^{-\int_{t_i}^{\tau_{i+1}} r(X(s)) \, \mathrm{d}s} V^{i+1}( X(\tau_{i+1})) \middle\vert X(t_i)=x\right]$ is a decreasing function and $E^{t_i}$ denotes the expectation under the $t_i$-forward adjusted measure.  Denote the probability measure under the $t_i$-forward adjusted measure as $\mathbb{Q}_{i}$.  $\mathbb{Q}_{i}$ and $\mathbb{P}$ are equivalent probability measures, so $\mathbb{P}[X_1(t) \leq X_2(t) \textrm{ for all } t\geq 0] =1$ implies that $\mathbb{Q}_{i}[X_1(t) \leq X_2(t) \textrm{ for all } t\geq 0] =1$.  Then $E^{t_i}\left[g(X(t_i)) \middle\vert X(\tau_i)=x\right]$ is a decreasing function of $x$.  At $x=x_i$, 
$$KP(\delta,x)=P(\delta,x)E^{t_i}\left[g(X(t_i))\middle\vert X(\tau_i)=x\right]$$
Then $K=E^{t_i}\left[g(X(t_i))|X(\tau_i)=x^i\right]$, so $K>E^{t_i}\left[g(X(t_i))\middle\vert X(\tau_i)=x\right]$ for all $x>x_i$.  Then $K P(\delta,x) > C^i(x)$ for all $x>x_i$.
\end{proof}

\begin{proof}[Proof of Proposition \ref{cirpropoab}]  
For Laguerre polynomials, the forward shift property is 
$$\frac{\mathrm{d}}{\mathrm{d}x} L_n^{(\alpha)} = -L_{n-1}^{(\alpha+1)}(x)$$
and the backward shift property is 
$$\frac{\mathrm{d}}{\mathrm{d}x}\left[ e^{-x} x^\alpha L_n^{(\alpha)}(x)\right] = (n+1) e^{-x} x^{\alpha-1} L_{n+1}^{(\alpha-1)}(x).$$

To derive the recursion for $a_{n,m}^{(\alpha)}(x)$, we first apply the backward shift, integrate by parts and then apply the forward shift.
\begin{eqnarray}
a_{n+1,m+1}^{(\alpha)} (x) &= & \int_0^x L_{n+1}^{(\alpha)}(y)L_{m+1}^{(\alpha)}(y)e^{-y}y^\alpha \,\mathrm{d}y \nonumber \\
&=& \frac{1}{m+1} \int_0^x L_{n+1}^{(\alpha)}(y) \, \mathrm{d}\left( e^{-y}y^{\alpha+1}L_{m}^{(\alpha+1)}(y) \right) \nonumber\\
&=& \frac{1}{m+1} \left[\left. L_{n+1}^{(\alpha)}(y)L_{m}^{(\alpha+1)}(y)e^{-y}y^{\alpha+1}\right|_0^x-\int_0^x e^{-y}y^{\alpha+1}L_{m}^{(\alpha+1)}(y) \, \mathrm{d}\left(L_{n+1}^{(\alpha)}(y) \right) \right] \nonumber\\
&=& \frac{1}{m+1} \left[ L_{n+1}^{(\alpha)}(x)L_{m}^{(\alpha+1)}(x)e^{-x}x^{\alpha+1} +\int_0^x e^{-y} y^{\alpha+1} L_{m}^{(\alpha+1)}(x)L_{n}^{(\alpha+1)}(x) \, \mathrm{d}y \right] \nonumber\\
&=& \frac{1}{m+1} \left[ L_{n+1}^{(\alpha)}(x)L_{m}^{(\alpha+1)}(x)e^{-x}x^{\alpha+1} + a_{n,m}^{(\alpha+1)}(x) \right] \label{alasteqn}
\end{eqnarray}

For $n\geq 1$, $m\geq 1$, $m\neq n$, by solving the above equation for $a_{n,m}^{\alpha+1}(x)$ and equating the expressions for $a_{n,m}^{\alpha+1}(x)$ and $a_{m,n}^{\alpha+1}(x)$, we have
\begin{equation*} 
a_{m,n}^{(\alpha)}(x) = \frac{e^{-x} x^{\alpha+1}}{m-n}\left( L_{n}^{(\alpha)}(x)L_{m-1}^{(\alpha+1)}(x)-L_{m}^{(\alpha)}(x)L_{n-1}^{(\alpha+1)}(x) \right).
 \end{equation*}

For $n\geq 1$, using the backward shift property, we have
\begin{equation*}
a_{0,n}(x)=\frac{1}{n} e^{-x} x^{\alpha+1} L_{n-1}^{(\alpha+1)}(x).
\end{equation*}

For $n=m$, we get the following from \eqref{alasteqn}:
\begin{equation*}
a_{n,n}^{(\alpha)}(x)=\frac{1}{n} \left[ L_n^{(\alpha)}(x) L_{n-1}^{(\alpha+1)}(x)e^{-x} x^{\alpha+1}+a_{n-1,n-1}^{(\alpha+1)}(x) \right], \; (n\geq 1), \quad a_{0,0}^{(\alpha)} = \gamma(\alpha+1,x),
 \end{equation*}
where $\gamma (\alpha+1,x)=\int_0^x e^{-y}y^\alpha \, \mathrm{d}y$ is the lower incomplete gamma function.

The coefficients $b_n^{(\alpha)}(x)$ are computed from equations \eqref{cirb1} and \eqref{cirb2} below.

For $n\geq 1$,
\begin{eqnarray} 
b_n^{(\alpha)}(x) &=& \int_0^x y^\alpha e^{-sy} L_n^{(\alpha)}(y) \, \mathrm{d}y \nonumber \\
 &=& \int_0^x e^{-(s-1)y}\frac{1}{n} \, \mathrm{d} \left[ e^{-y} y^{\alpha+1} L_{n-1}^{(\alpha+1)}(y) \right] \nonumber\\
 &=& \left. \frac{1}{n}e^{-sy}y^{\alpha+1} L_{n-1}^{(\alpha+1)} (y) \right|_0^x +\frac{s-1}{n}\int_0^x e^{-sy} y^{\alpha+1}L_{n-1}^{(\alpha+1)}(y) \,\mathrm{d}y \nonumber\\
 &=& \frac{1}{n}e^{-sx}x^{\alpha+1} L_{n-1}^{(\alpha+1)}(x) + \frac{s-1}{n} b_{n-1}^{(\alpha+1)}(x) \label{cirb1}
 \end{eqnarray}
 
 \begin{eqnarray} b_0^{(\alpha)}(x)& =& \int_0^x y^\alpha e^{-sy} L_0^{(\alpha)}(y) \,\mathrm{d}y \nonumber \\
 &=& \frac{1}{s^{\alpha+1}} \gamma (\alpha+1,sx) \label{cirb2}
 \end{eqnarray}
 \end{proof}

\begin{proof}[Proof of Proposition \ref{vasicekpropoab}] 
For Hermite polynomials, the forward shift property is 
$$\frac{\mathrm{d}}{\mathrm{d}x} H_n(x) = 2nH_{n-1}(x)$$
and the backward shift property is 
$$\frac{\mathrm{d}}{\mathrm{d}x}\left[ e^{-x^2}H_n(x)\right] = -e^{-x^2}H_{n+1}(x).$$
 
To derive the recursion for $a_{n,m}(x)$, we first apply the backward shift, integrate by parts and then apply the forward shift.
\begin{eqnarray}
a_{n+1,m+1}(x) &=& \int_{-\infty}^x H_{n+1}(y)H_{m+1}(y) e^{-y^2} \, \mathrm{d}y  \nonumber \\
&=& -\int_{-\infty}^x H_{n+1}(y) \,\mathrm{d}\left(e^{-y^2}H_m(y) \right) \nonumber \\
&=& \left. -H_{n+1}(y)H_m(y)e^{-y^2}\right|_{-\infty}^x + \int_{-\infty}^x e^{-y^2} H_m(y) \,\mathrm{d}\left(H_{n+1}(y) \right)  \nonumber \\
&=& -H_{n+1}(x)H_m(x)e^{-x^2}+ \int_{-\infty}^x e^{-y^2} H_m(y) 2(n+1)H_n(y) \, \mathrm{d}y \nonumber \\
&=& -H_{n+1}(x)H_m(x)e^{-x^2}+2(n+1)a_{n,m}(x) \label{ahrecur1}
\end{eqnarray}
 
 Exchanging the roles of $n$ and $m$, we have 
 \begin{equation} \label{ahrecur2} a_{m+1,n+1}(x) = -H_{m+1}(x)H_n(x)e^{-x^2} +2(m+1)a_{m,n}(x).\end{equation}
 If $m\neq n$, then subtracting \eqref{ahrecur1} from \eqref{ahrecur2} and rearranging, we have 
 \begin{equation*} a_{n,m}(x) = \frac{H_n(x)H_{m+1}(x)-H_m(x)H_{n+1}(x)}{2(m-n)}e^{-x^2}.  \end{equation*}
 
 $a_{n,n}(x)$ can be computed recursively as follows:
 \begin{equation} a_{n,n}(x) = -H_{n-1}(x)H_n(x) e^{-x^2} +2na_{n-1,n-1}(x), \; (n\geq 1),\quad a_{0,0}(x)=\sqrt{\pi}\Phi(\sqrt{2}x), \nonumber \end{equation}
where $\Phi(x)$ is the cumulative distribution function of a standard normal distribution.

For $n\geq1$,
\begin{eqnarray}
b_n(s,x) &=& \int_{-\infty}^{x} e^{su-u^2} H_n(u) \, \mathrm{d}u \nonumber \\
&=& -\int_{-\infty}^x e^{su} \, \mathrm{d}\left[e^{-u^2} H_{n-1}(u) \right] \nonumber \\
&=&\left. -e^{su-u^2} H_{n-1}(u) \right|_{-\infty}^x +\int_{-\infty}^x s e^{su-u^2} H_{n-1}(u) \, \mathrm{d}u \nonumber \\
&=& -e^{sx-x^2} H_{n-1} (x) +sb_{n-1}(s,x) \nonumber
\end{eqnarray}

\begin{equation} 
b_0(s,x)=\frac{1}{2} e^{\frac{s^2}{4}} \sqrt{\pi}(\mathrm{Erf}(\frac{1}{2}(2x-s))+1), \nonumber
\end{equation}
where $\mathrm{Erf}(x)$ is the error function.
\end{proof}

\setlength{\bibsep}{0pt}
\bibliographystyle{model2-names}
\bibliography{biblio}

\begin{thebibliography}{46}
\expandafter\ifx\csname natexlab\endcsname\relax\def\natexlab#1{#1}\fi
\expandafter\ifx\csname url\endcsname\relax
  \def\url#1{\texttt{#1}}\fi
\expandafter\ifx\csname urlprefix\endcsname\relax\def\urlprefix{URL }\fi
\providecommand{\eprint}[2][]{\url{#2}}
\providecommand{\bibinfo}[2]{#2}
\ifx\xfnm\relax \def\xfnm[#1]{\unskip,\space#1}\fi
\bibitem[{Ahn and Gao(1999)}]{ahngao}
\bibinfo{author}{Ahn, {\relax D.-H}.}, \bibinfo{author}{Gao, B.},
  \bibinfo{year}{1999}.
\newblock \bibinfo{title}{A parametric nonlinear model of term structure
  dynamics}.
\newblock \bibinfo{journal}{Review of Financial Studies} \bibinfo{volume}{12},
  \bibinfo{pages}{721--762}.
\bibitem[{Albanese and Kuznetsov(2004)}]{alba}
\bibinfo{author}{Albanese, C.}, \bibinfo{author}{Kuznetsov, A.},
  \bibinfo{year}{2004}.
\newblock \bibinfo{title}{Unifying the three volatility models}.
\newblock \bibinfo{journal}{RISK} \bibinfo{volume}{17},
  \bibinfo{pages}{94--98}.
\bibitem[{Albeverio and {R\"{u}diger}(2003)}]{albev}
\bibinfo{author}{Albeverio, S.}, \bibinfo{author}{{R\"{u}diger}, B.},
  \bibinfo{year}{2003}.
\newblock \bibinfo{title}{Infinite-dimensional stochastic differential
  equations obtained by subordination and related {D}irichlet forms}.
\newblock \bibinfo{journal}{Journal of Functional Analysis}
  \bibinfo{volume}{204}, \bibinfo{pages}{122--156}.
\bibitem[{Barndorff-Nielsen(1998)}]{barndorff}
\bibinfo{author}{Barndorff-Nielsen, O.E.}, \bibinfo{year}{1998}.
\newblock \bibinfo{title}{Processes of normal inverse {G}aussian type}.
\newblock \bibinfo{journal}{Finance and Stochastics} \bibinfo{volume}{2},
  \bibinfo{pages}{41--68}.
\bibitem[{Barndorff-Nielsen and Levendorski\v{i}(2001)}]{barnlev}
\bibinfo{author}{Barndorff-Nielsen, O.E.}, \bibinfo{author}{Levendorski\v{i},
  S.}, \bibinfo{year}{2001}.
\newblock \bibinfo{title}{Feller processes of normal inverse {Gaussian} type}.
\newblock \bibinfo{journal}{Quantitative Finance} \bibinfo{volume}{1},
  \bibinfo{pages}{318--331}.
\bibitem[{Beaglehole and Tenney(1992)}]{beagle}
\bibinfo{author}{Beaglehole, D.R.}, \bibinfo{author}{Tenney, M.},
  \bibinfo{year}{1992}.
\newblock \bibinfo{title}{A non-linear equilibrium model of the term structure
  of interest rates: Corrections and additions}.
\newblock \bibinfo{journal}{Journal of Financial Economics}
  \bibinfo{volume}{32}, \bibinfo{pages}{345--353}.
\bibitem[{Ben-Ameur et~al.(2007)Ben-Ameur, Breton, Karoui and L'Ecuyer}]{bbkl}
\bibinfo{author}{Ben-Ameur, H.}, \bibinfo{author}{Breton, M.},
  \bibinfo{author}{Karoui, L.}, \bibinfo{author}{L'Ecuyer, P.},
  \bibinfo{year}{2007}.
\newblock \bibinfo{title}{A dynamic programming approach for pricing options
  embedded in bonds}.
\newblock \bibinfo{journal}{Journal of Economic Dynamics \& Control}
  \bibinfo{volume}{31}, \bibinfo{pages}{2212--2233}.
\bibitem[{Bochner(1949)}]{bochner}
\bibinfo{author}{Bochner, S.}, \bibinfo{year}{1949}.
\newblock \bibinfo{title}{Diffusion equation and stochastic processes}.
\newblock \bibinfo{journal}{Proceedings of the National Academy of Sciences of
  the United States of America} \bibinfo{volume}{35},
  \bibinfo{pages}{368--370}.
\bibitem[{Bochner(1955)}]{bochner2}
\bibinfo{author}{Bochner, S.}, \bibinfo{year}{1955}.
\newblock \bibinfo{title}{Harmonic Analysis and the Theory of Probability}.
\newblock \bibinfo{publisher}{Univ. of California Press}.
\bibitem[{Borodin and Salminen(2002)}]{boro}
\bibinfo{author}{Borodin, A.N.}, \bibinfo{author}{Salminen, P.},
  \bibinfo{year}{2002}.
\newblock \bibinfo{title}{Handbook of {B}rownian {M}otion}.
\newblock \bibinfo{publisher}{{Birkh\"{a}user}}, \bibinfo{address}{Boston, MA}.
  \bibinfo{edition}{2nd} edition.
\bibitem[{Boyarchenko and Levendorski\v{i}(2006)}]{boyar}
\bibinfo{author}{Boyarchenko, N.}, \bibinfo{author}{Levendorski\v{i}, S.},
  \bibinfo{year}{2006}.
\newblock \bibinfo{title}{The eigenfunction expansion method in multi-factor
  quadratic term structure models}.
\newblock \bibinfo{journal}{Mathematical Finance} \bibinfo{volume}{17},
  \bibinfo{pages}{503--539}.
\bibitem[{Brennan and Schwartz(1977)}]{brennan}
\bibinfo{author}{Brennan, M.J.}, \bibinfo{author}{Schwartz, E.S.},
  \bibinfo{year}{1977}.
\newblock \bibinfo{title}{Savings bonds, retractable bonds and callable bonds}.
\newblock \bibinfo{journal}{Journal of Financial Economics}
  \bibinfo{volume}{5}, \bibinfo{pages}{67--88}.
\bibitem[{Brigo and Mercurio(2001)}]{brigo2}
\bibinfo{author}{Brigo, D.}, \bibinfo{author}{Mercurio, F.},
  \bibinfo{year}{2001}.
\newblock \bibinfo{title}{On deterministic-shift extensions of short-rate
  models}.
\newblock \bibinfo{journal}{Finance and Stochastics} \bibinfo{volume}{5},
  \bibinfo{pages}{369--387}.
\bibitem[{{B\"{u}ttler}(1995)}]{buttler}
\bibinfo{author}{{B\"{u}ttler}, H.}, \bibinfo{year}{1995}.
\newblock \bibinfo{title}{Evaluation of callable bonds: {F}inite difference
  methods, stability and accuracy}.
\newblock \bibinfo{journal}{Economic Journal} \bibinfo{volume}{105},
  \bibinfo{pages}{374--384}.
\bibitem[{{B\"{u}ttler} and Waldvogel(1996)}]{bw}
\bibinfo{author}{{B\"{u}ttler}, H.}, \bibinfo{author}{Waldvogel, J.},
  \bibinfo{year}{1996}.
\newblock \bibinfo{title}{Pricing callable bonds by means of {G}reen's
  function}.
\newblock \bibinfo{journal}{Mathematical Finance} \bibinfo{volume}{6},
  \bibinfo{pages}{53--88}.
\bibitem[{Chen(2005)}]{chen}
\bibinfo{author}{Chen, Z.Q.}, \bibinfo{year}{2005}.
\newblock \bibinfo{title}{On {F}eynman-{K}ac perturbation of symmetric {M}arkov
  processes}, in: \bibinfo{booktitle}{Proceedings of Functional Analysis IX},
  pp. \bibinfo{pages}{39--43}.
\bibitem[{Chen and Song(2005)}]{chensong}
\bibinfo{author}{Chen, Z.Q.}, \bibinfo{author}{Song, R.}, \bibinfo{year}{2005}.
\newblock \bibinfo{title}{Two-sided eigenvalue estimates for subordinate
  processes in domains}.
\newblock \bibinfo{journal}{Journal of Functional Analysis}
  \bibinfo{volume}{226}, \bibinfo{pages}{90--113}.
\bibitem[{Cont and Tankov(2004)}]{cont}
\bibinfo{author}{Cont, R.}, \bibinfo{author}{Tankov, P.}, \bibinfo{year}{2004}.
\newblock \bibinfo{title}{Financial Modelling with Jump Processes}.
\newblock \bibinfo{publisher}{Chapman \& Hall/CRC Press}.
\bibitem[{Cox et~al.(1985)Cox, Ingersoll and Ross}]{cir}
\bibinfo{author}{Cox, J.C.}, \bibinfo{author}{Ingersoll, J.E.},
  \bibinfo{author}{Ross, S.A.}, \bibinfo{year}{1985}.
\newblock \bibinfo{title}{A theory of the term structure of interest rates}.
\newblock \bibinfo{journal}{Econometrica} \bibinfo{volume}{53},
  \bibinfo{pages}{385--408}.
\bibitem[{Davies(2007)}]{davies}
\bibinfo{author}{Davies, E.B.}, \bibinfo{year}{2007}.
\newblock \bibinfo{title}{Linear Operators and their Spectra}.
\newblock \bibinfo{publisher}{Cambridge University Press},
  \bibinfo{address}{Cambridge, UK}.
\bibitem[{Davydov and Linetsky(2003)}]{davy}
\bibinfo{author}{Davydov, D.}, \bibinfo{author}{Linetsky, V.},
  \bibinfo{year}{2003}.
\newblock \bibinfo{title}{Pricing options on scalar diffusions: An
  eigenfunction expansion approach}.
\newblock \bibinfo{journal}{Operations Research} \bibinfo{volume}{51},
  \bibinfo{pages}{185--209}.
\bibitem[{d'Halluin et~al.(2001)d'Halluin, Forsyth, Vetzal and Labahn}]{dfvl}
\bibinfo{author}{d'Halluin, Y.}, \bibinfo{author}{Forsyth, P.A.},
  \bibinfo{author}{Vetzal, K.R.}, \bibinfo{author}{Labahn, G.},
  \bibinfo{year}{2001}.
\newblock \bibinfo{title}{A numerical {PDE} approach for pricing callable
  bonds}.
\newblock \bibinfo{journal}{Applied Mathematical Finance} \bibinfo{volume}{8},
  \bibinfo{pages}{49--77}.
\bibitem[{Dynkin(1969)}]{dynkin}
\bibinfo{author}{Dynkin, E.B.}, \bibinfo{year}{1969}.
\newblock \bibinfo{title}{Game variant of a problem on optimal stopping}.
\newblock \bibinfo{journal}{Soviet Mathematics-Doklady} \bibinfo{volume}{10},
  \bibinfo{pages}{270--274}.
\bibitem[{Farto and {V\'{a}zquez}(2005)}]{farto}
\bibinfo{author}{Farto, J.}, \bibinfo{author}{{V\'{a}zquez}, C.},
  \bibinfo{year}{2005}.
\newblock \bibinfo{title}{Numerical techniques for pricing callable bonds with
  notice}.
\newblock \bibinfo{journal}{Applied Mathematics and Computation}
  \bibinfo{volume}{161}, \bibinfo{pages}{989--1013}.
\bibitem[{de~Frutos(2008)}]{frutos}
\bibinfo{author}{de~Frutos, J.}, \bibinfo{year}{2008}.
\newblock \bibinfo{title}{A spectral method for bonds}.
\newblock \bibinfo{journal}{Computers \& Operations Research}
  \bibinfo{volume}{35}, \bibinfo{pages}{64--75}.
\bibitem[{Gorovoi and Linetsky(2004)}]{gorovoi}
\bibinfo{author}{Gorovoi, V.}, \bibinfo{author}{Linetsky, V.},
  \bibinfo{year}{2004}.
\newblock \bibinfo{title}{Black's model of interest rates as options,
  eigenfunction expansions and {J}apanese interest rates}.
\newblock \bibinfo{journal}{Mathematical Finance} \bibinfo{volume}{14},
  \bibinfo{pages}{49--78}.
\bibitem[{Gorovoi and Linetsky(2007)}]{gorovoi2}
\bibinfo{author}{Gorovoi, V.}, \bibinfo{author}{Linetsky, V.},
  \bibinfo{year}{2007}.
\newblock \bibinfo{title}{Intensity-based valuation of residential mortgages:
  An analytically tractable model}.
\newblock \bibinfo{journal}{Mathematical Finance} \bibinfo{volume}{17},
  \bibinfo{pages}{541--573}.
\bibitem[{Gradshteyn and Ryzhik(2007)}]{grad}
\bibinfo{author}{Gradshteyn, I.S.}, \bibinfo{author}{Ryzhik, I.},
  \bibinfo{year}{2007}.
\newblock \bibinfo{title}{Tables of Integrals, Series and Products}.
\newblock \bibinfo{publisher}{Academic Press}, \bibinfo{address}{New York}.
  \bibinfo{edition}{7th} edition.
\bibitem[{Ikeda and Watanabe(1977)}]{ikeda}
\bibinfo{author}{Ikeda, N.}, \bibinfo{author}{Watanabe, S.},
  \bibinfo{year}{1977}.
\newblock \bibinfo{title}{A comparison theorem for solutions of stochastic
  differential equations and its applications}.
\newblock \bibinfo{journal}{Osaka Journal of Mathematics} \bibinfo{volume}{14},
  \bibinfo{pages}{619--633}.
\bibitem[{Ito and McKean(1974)}]{itomckean}
\bibinfo{author}{Ito, K.}, \bibinfo{author}{McKean, H.}, \bibinfo{year}{1974}.
\newblock \bibinfo{title}{Diffusion Processes and their Sample Paths}.
\newblock \bibinfo{publisher}{Springer}, \bibinfo{address}{Berlin}.
\bibitem[{Karlin and Taylor(1981)}]{karlin}
\bibinfo{author}{Karlin, S.}, \bibinfo{author}{Taylor, H.M.},
  \bibinfo{year}{1981}.
\newblock \bibinfo{title}{A {S}econd {C}ourse in {S}tochastic {P}rocesses}.
\newblock \bibinfo{publisher}{Academic Press}, \bibinfo{address}{San Diego,
  CA}.
\bibitem[{Lebedev(1965)}]{lebedev}
\bibinfo{author}{Lebedev, N.N.}, \bibinfo{year}{1965}.
\newblock \bibinfo{title}{Special {F}unctions and {T}heir {A}pplications}.
\newblock \bibinfo{publisher}{Prentice-Hall, Inc.}, \bibinfo{address}{Englewood
  Cliffs, N.J.}
\bibitem[{Leippold and Wu(2002)}]{leip}
\bibinfo{author}{Leippold, M.}, \bibinfo{author}{Wu, L.}, \bibinfo{year}{2002}.
\newblock \bibinfo{title}{Asset pricing under the quadratic class}.
\newblock \bibinfo{journal}{Journal of Financial and Quantitative Analysis}
  \bibinfo{volume}{37}, \bibinfo{pages}{271--295}.
\bibitem[{Lewis(1998)}]{lewis1}
\bibinfo{author}{Lewis, A.}, \bibinfo{year}{1998}.
\newblock \bibinfo{title}{Applications of eigenfunction expansions in
  continuous-time finance}.
\newblock \bibinfo{journal}{Mathematical Finance} \bibinfo{volume}{8},
  \bibinfo{pages}{349--383}.
\bibitem[{Li and Linetsky(2011)}]{lingfei2}
\bibinfo{author}{Li, L.}, \bibinfo{author}{Linetsky, V.}, \bibinfo{year}{2011}.
\newblock \bibinfo{title}{Time-changed {O}rnstein-{U}hlenbeck processes and
  their applications in commodity derivative models}.
\newblock \bibinfo{note}{To appear in Mathematical Finance}.
\bibitem[{Linetsky(2004)}]{linet1}
\bibinfo{author}{Linetsky, V.}, \bibinfo{year}{2004}.
\newblock \bibinfo{title}{The spectral decomposition of the option value}.
\newblock \bibinfo{journal}{International Journal of Theoretical and Applied
  Finance} \bibinfo{volume}{7}, \bibinfo{pages}{337--384}.
\bibitem[{Linetsky(2008)}]{linethandbook}
\bibinfo{author}{Linetsky, V.}, \bibinfo{year}{2008}.
\newblock \bibinfo{title}{Spectral methods in derivatives pricing}, in:
  \bibinfo{editor}{Birge, J.}, \bibinfo{editor}{Linetsky, V.} (Eds.),
  \bibinfo{booktitle}{Handbooks in Operations Research and Management Science}.
  \bibinfo{publisher}{Elsevier}, \bibinfo{address}{Amsterdam}.
  volume~\bibinfo{volume}{15}. chapter~\bibinfo{chapter}{6}, pp.
  \bibinfo{pages}{223--299}.
\bibitem[{Madan et~al.(1998)Madan, Carr and Chang}]{madan}
\bibinfo{author}{Madan, D.B.}, \bibinfo{author}{Carr, P.P.},
  \bibinfo{author}{Chang, E.C.}, \bibinfo{year}{1998}.
\newblock \bibinfo{title}{The {V}ariance {G}amma process and option pricing}.
\newblock \bibinfo{journal}{European Finance Review} \bibinfo{volume}{2},
  \bibinfo{pages}{79--105}.
\bibitem[{McKean(1956)}]{mckean}
\bibinfo{author}{McKean, H.}, \bibinfo{year}{1956}.
\newblock \bibinfo{title}{Elementary solutions for certain parabolic partial
  differential equations}.
\newblock \bibinfo{journal}{Transactions of the American Mathematical Society}
  \bibinfo{volume}{82}, \bibinfo{pages}{519--548}.
\bibitem[{Mendoza-Arriaga et~al.(2010)Mendoza-Arriaga, Carr and
  Linetsky}]{mendoza}
\bibinfo{author}{Mendoza-Arriaga, R.}, \bibinfo{author}{Carr, P.},
  \bibinfo{author}{Linetsky, V.}, \bibinfo{year}{2010}.
\newblock \bibinfo{title}{Time-changed {M}arkov processes in unified
  credit-equity modeling}.
\newblock \bibinfo{journal}{Mathematical Finance} \bibinfo{volume}{20},
  \bibinfo{pages}{527--569}.
\bibitem[{Mendoza-Arriaga and Linetsky(2011)}]{mendoza2}
\bibinfo{author}{Mendoza-Arriaga, R.}, \bibinfo{author}{Linetsky, V.},
  \bibinfo{year}{2011}.
\newblock \bibinfo{title}{Constructing {M}arkov processes with dependent jumps
  by multivariate subordination: {a}pplications to multi-name credit-equity
  modeling}.
\newblock \bibinfo{note}{Preprint}.
\bibitem[{Nikiforov and Uvarov(1988)}]{niki}
\bibinfo{author}{Nikiforov, A.F.}, \bibinfo{author}{Uvarov, V.B.},
  \bibinfo{year}{1988}.
\newblock \bibinfo{title}{Special Functions of Mathematical Physics: A Unified
  Introduction with Applications}.
\newblock \bibinfo{publisher}{{Birkh\"{a}user}}.
\bibitem[{Phillips(1952)}]{phillip}
\bibinfo{author}{Phillips, R.S.}, \bibinfo{year}{1952}.
\newblock \bibinfo{title}{On the generation of semigroups of linear operators}.
\newblock \bibinfo{journal}{Pacific Journal of Mathematics}
  \bibinfo{volume}{2}, \bibinfo{pages}{343--369}.
\bibitem[{Sato(1999)}]{sato}
\bibinfo{author}{Sato, K.}, \bibinfo{year}{1999}.
\newblock \bibinfo{title}{{L\'{e}vy} {P}rocesses and {I}nfinitely {D}ivisible
  {D}istributions}.
\newblock \bibinfo{publisher}{Cambridge Univ. Press},
  \bibinfo{address}{Cambridge}.
\bibitem[{Schilling et~al.(2010)Schilling, Song and {Vondra\v{c}ek}}]{schill}
\bibinfo{author}{Schilling, R.L.}, \bibinfo{author}{Song, R.},
  \bibinfo{author}{{Vondra\v{c}ek}, Z.}, \bibinfo{year}{2010}.
\newblock \bibinfo{title}{Bernstein Functions: Theory and Applications}.
\newblock de Gruyter Studies in Mathematics 37, \bibinfo{publisher}{Walter de
  Gruyter}, \bibinfo{address}{Berlin, Germany}.
\bibitem[{Vasicek(1977)}]{vasicek}
\bibinfo{author}{Vasicek, O.A.}, \bibinfo{year}{1977}.
\newblock \bibinfo{title}{An equilibrium characterization of the term
  structure}.
\newblock \bibinfo{journal}{Journal of Financial Economics}
  \bibinfo{volume}{5}, \bibinfo{pages}{177--188}.

\end{thebibliography}

\begin{table}[!ht]
\caption{Call prices}
\label{callprice}
\footnotesize
\begin{tabularx}{\textwidth} {X X }
\toprule
Exercise date & Call Price\\
\midrule
$t_{11}=10.172$ & 1.025 \\
$t_{12}=11.172$ & 1.020 \\
$t_{13}=12.172$ & 1.015 \\
$t_{14}=13.172$ & 1.010 \\
$t_{15}=14.172$ & 1.005 \\
$t_{16}=10.172$ to $t_{20}=19.172$ & 1.000 \\
\bottomrule
\end{tabularx}
 \end{table}

 \begin{table}[!ht]
\caption{Parameter values}
\label{parameters}
\footnotesize
\begin{tabularx}{\textwidth} {X X X}
\toprule
& Vasicek & CIR\\
\midrule
$\kappa$ & 0.44178462& 0.14294371 \\
$\sigma$ & 0.13264223& 0.38757496 \\
$\theta$ & 0.098397028& 0.133976855 \\
\bottomrule
\end{tabularx}
 \end{table}
 
\begin{table}[!ht]
\caption{Break-even short rates}
\label{breakeven1}
\footnotesize
\begin{tabularx}{\textwidth} {l X X X X X X}
\toprule
Time& CIR & Vasicek  & SubCIR, JD& SubCIR, PJ& SubVasicek, JD & SubVasicek, PJ \\
\midrule
$\tau_{20}$ & 0.03388791& 0.02706597&0.03614163&0.03672670& 0.03189678& 0.03348832\\
$\tau_{19}$ & 0.01792789& -0.01012520 &0.02292836&0.02439808& 0.00299207& 0.00734621\\
$\tau_{18}$ & 0.00978966& -0.03655983&0.01665424&0.01758017&-0.01809927&-0.01208475\\
$\tau_{17}$ & 0.00488209& -0.05701483 &0.01161351&0.01333251&-0.03477951&-0.02766935\\
$\tau_{16}$ & 0.00157881& -0.07350682&0.00873978&0.01047766&-0.04847549&-0.04061315\\
$\tau_{15}$ & n.a. & -0.09100438&n.a.&n.a.&-0.06370872&-0.05539452\\
$\tau_{14}$ & n.a. & -0.10481935&n.a.&n.a.&-0.07568237&-0.06698556\\
$\tau_{13}$ & n.a. & -0.11653925&n.a.&n.a.&-0.08590952&-0.07694429\\
$\tau_{12}$ & n.a. & -0.12671317&n.a.&n.a.&-0.09485232&-0.08570132\\
$\tau_{11}$ & n.a. & -0.13566906&n.a.&n.a.&-0.10277749&-0.09350086\\
\bottomrule
\end{tabularx}
\end{table}

  \begin{table}[!ht]
\caption{Convergence and CPU times for pricing the callable bond with dynamic truncation level for initial short rate $r=0.05$}
\label{conv1}
\footnotesize
\begin{tabularx}{\textwidth} {lXXc}
\toprule
Pricing Error & Average N at $\tau_{20}$, ..., $\tau_{11}$ and $t_0$ & Maximum $N$ at $\tau_{20}$, ..., $\tau_{11}$ and $t_0$ & CPU time (ms)\\
\midrule
\multicolumn{4}{c}{CIR}\\
\midrule
$10^{-5}$&  6.0, 3.4, 3.2, 3.1, 3.0, 4.0, 4.0, 4.0, 4.0, 4.0, 2 &6, 5, 5, 5, 4, 4, 4, 4, 4, 4, 2 & 1.1 \\
$10^{-6}$&8.9, 7.0, 5.3, 5.4, 5.1, 5.0, 6.0, 6.0, 6.0, 6.0, 2&9, 8, 8, 8, 7, 5, 6, 6, 6, 6, 2 &  1.4\\
$10^{-7}$&10.9, 11.0, 7.9, 9.0, 7.0, 7.0, 7.0, 7.0, 7.0, 3 &11, 11, 12, 11, 11, 7, 7, 7, 7, 7, 3 & 1.9 \\
\midrule
\multicolumn{4}{c}{Vasicek}\\
\midrule
$10^{-5}$ & 4.2, 5.8, 6.0, 4.3, 5.8, 5.9, 5.2, 5.2, 5.2, 5.2, 2 &5, 6, 6, 6, 6, 6, 7, 7, 7, 7, 2 & 0.8\\
$10^{-6}$& 6.0, 8.3, 9.8, 9.8, 9.2, 9.0, 9.0, 9.3, 9.9, 10.0, 3&6, 10, 10, 10, 11, 10, 11, 11, 11, 11, 3&1.3 \\
$10^{-7}$&6.1, 12.0, 12.0, 13.0, 13.0, 12.8, 12.9, 13.9, 13.8, 13.5, 3& 7, 13, 14, 14, 14, 13, 13, 14, 14, 14, 3&1.8  \\
\midrule
\multicolumn{4}{c}{SubCIR, Jump-diffusion}\\
\midrule
$10^{-5}$& 10.0, 9.9, 9.0, 5.9, 7.0, 7.0, 7.0, 6.0, 6.0, 6.0, 3 & 10, 10, 11, 11, 10, 7, 7, 6, 6, 6, 3& 2.1 \\
$10^{-6}$&11.9, 12.8, 13.4, 8.6, 9.7, 10., 8.0, 8.0, 8.0, 8.0, 3&12, 13, 14, 14, 14, 10, 8, 8, 8, 8, 3& 2.7\\
$10^{-7}$&14.0, 15.8, 18.6, 19.2, 13.5, 16.0, 11.0, 10.0, 10.0, 10.0, 4& 14, 16, 19, 21, 20, 16, 11, 10, 10, 10, 4&  4.0\\
\midrule
\multicolumn{4}{c}{SubCIR, Pure jump}\\
\midrule
$10^{-5}$& 10.9, 10.8, 9.0, 5.7, 5.6, 5.0, 6.0, 6.0, 6.0, 6.0, 3& 11, 11, 12, 12, 11, 5, 6, 6, 6, 6, 3& 2.2 \\
$10^{-6}$& 12.9, 14.8, 18.3, 16.7, 10.9, 13.0, 9.0, 8.0, 8.0, 8.0, 3& 13, 15, 19, 20, 20, 13, 9, 8, 8, 8, 3& 3.6\\
$10^{-7}$&16.1, 27.7, 27.7, 35.0, 22.3, 31.0, 13.0, 12.0, 12.0, 12.0, 4& 17, 35, 28, 36, 39, 31, 13, 12, 12, 12, 4& 8.7 \\
\midrule
\multicolumn{4}{c}{SubVasicek, Jump-diffusion}\\
\midrule
$10^{-5}$&  6.0, 6.3, 6.3, 7.9, 8.0, 7.3, 7.1, 7.2, 7.1, 7.1, 3& 6, 8, 10, 8, 8, 9, 8, 9, 9, 9, 3& 1.5 \\
$10^{-6}$& 6.0, 12.0, 12.2, 13.2, 13.2, 13.2, 13.0, 13.9, 14.0, 14.0, 3&7, 13, 14, 15, 15, 15, 15, 15, 15, 14, 3& 2.5\\
$10^{-7}$& 8.0, 18.3, 19.6, 20.6, 20.9, 21.0, 20.2, 20.0, 21.6, 21.5, 4& 8, 21, 22, 21, 23, 23, 22, 22, 22, 22, 4&  4.2\\
\midrule
\multicolumn{4}{c}{SubVasicek, Pure jump}\\
\midrule
$10^{-5}$& 6.0, 10.3, 11.8, 12.3, 13.1, 13.0, 12.9, 12.6, 13.7, 13.8, 3  &6, 12, 14, 15, 16, 15, 15, 15, 15, 15, 3& 2.5\\
$10^{-6}$& 7.0, 22.3, 23.4, 25.5, 26.8, 27.8, 27.6, 27.6, 28.7, 28.5, 4&8, 29, 30, 29, 29, 31, 32, 30, 30, 31, 4& 6.2\\
$10^{-7}$& 8.0, 44.6, 43.7, 47.8, 48.6, 47.2, 51.0, 51.1, 54.0, 52.1, 5&8, 55, 52, 57, 53, 54, 55, 57, 57, 56, 5&  15.9\\
\bottomrule
\end{tabularx}
 \end{table}

      \begin{table}[!ht]
\caption{CIR model: Values of the callable bond for initial short rate $r$ obtained by five methods}
\label{cirsummary}
\footnotesize
\begin{tabularx}{\textwidth} {XXXXXX}
\toprule
$r$& BW & DFVL & BBKL & F & this paper\\
\midrule
0.01 & 0.9392&0.93926&0.93921&0.93922&0.939259\\
0.02 & 0.9159&0.91598&0.91595&0.91596&0.915992\\
0.03& 0.8933&0.89333&0.89330&0.89331&0.893341\\
0.04& 0.8712&0.87127&0.87125&0.87125&0.871290\\
0.05&0.8498&0.84980&0.84978&0.84979&0.849823\\
0.06& 0.8289&0.82890&0.82888&0.82889&0.828923\\
0.07&0.8085&0.80855&0.80854&0.80854&0.808577\\
0.08& 0.7887&0.78874&0.78873&0.78873&0.788769\\
0.09& 0.7694&0.76945&0.76945&0.76945&0.769484\\
0.10& 0.7507&0.75067&0.75067&0.75067&0.750708\\
\bottomrule
\end{tabularx}
\end{table}

      \begin{table}[!ht]
\caption{Vasicek model: Values of the callable bond for initial short rate $r$ obtained by four methods}
\label{vasiceksummary}
\footnotesize
\begin{tabularx}{\textwidth} {XXXXX}
\toprule
 $r$& BW & DFVL & BBKL & this paper\\
\midrule
0.01 & 0.8556&0.84282&0.84285&0.842845\\
0.02 & 0.8338&0.82627&0.82630&0.826294\\
0.03& 0.8223&0.81010&0.81009&0.810091\\
0.04& 0.8062&0.79420&0.79423&0.794230\\
0.05&0.7904&0.77868&0.77871&0.778702\\
0.06& 0.7749&0.76348&0.76351&0.763502\\
0.07&0.7598&0.74860&0.74862&0.748621\\
0.08& 0.7450&0.73403&0.73406&0.734053\\
0.09& 0.7305&0.71977&0.71980&0.719792\\
0.10& 0.7163&0.70578&0.70583&0.705830\\
\bottomrule
\end{tabularx}
 \end{table}
 
     \begin{table}[!ht]
\caption{Subordinated models: Values of the callable bond for initial short rate $r$ obtained by the eigenfunction expansion method}
\label{subcirsummary1}
\footnotesize
\begin{tabularx}{\textwidth} {XXXXX}
\toprule
 $r$& SubCIR, JD & SubCIR, PJ& SubVasicek, JD& SubVasicek, PJ\\
\midrule
0.01 & 0.967362&0.972668&0.874805&0.884935\\
0.02 & 0.941069&0.946130&0.855193&0.864408\\
0.03& 0.915446&0.920208& 0.835999&0.844285\\
0.04& 0.890481&0.894892& 0.817216&0.824562\\
0.05& 0.866160&0.870174 &0.798837&0.805233\\
0.06& 0.842470&0.846044& 0.780854&0.786293\\
0.07& 0.819396&0.822492 & 0.763261&0.767737\\
0.08& 0.796927&0.799510&0.746050&0.749559\\
0.09& 0.775050&0.777087&0.729215& 0.731754\\
0.10& 0.753752&0.755215&0.712749&0.714318\\
\bottomrule
\end{tabularx}
 \end{table}

\begin{table}[!ht]
\caption{Put prices}
\label{putprice}
\footnotesize
\begin{tabularx}{\textwidth} {X X }
\toprule
Exercise date & Put Price\\
\midrule
$t_{11}=10.172$ & 1.015 \\
$t_{12}=11.172$ & 1.010 \\
$t_{13}=12.172$ & 1.005 \\
$t_{14}=13.172$ & 1.000 \\
$t_{15}=14.172$ & 0.995 \\
$t_{16}=10.172$ to $t_{20}=19.172$ & 0.990 \\
\bottomrule
\end{tabularx}
 \end{table}

\begin{table}[!ht]
\caption{Break-even short rates for callable and putable bond}
\label{breakevencallput}
\footnotesize
\begin{tabularx}{\textwidth} {l X X X X X X}
\toprule
Time& CIR & Vasicek  & SubCIR, JD& SubCIR, PJ& SubVasicek, JD & SubVasicek, PJ \\
\midrule
\multicolumn{7}{c}{Call Option}\\
\midrule   
$\tau_{20}$ & 0.03388791& 0.02706597&0.03614163&0.03672670&0.03189678&0.03348832\\
$\tau_{19}$ & 0.03050674& 0.01653941&0.03271682&0.03328046&0.02560234&0.02738706\\
$\tau_{18}$ & 0.03032523& 0.01570707&0.03248071&0.03298851&0.02523355&0.02696967\\
$\tau_{17}$ & 0.03031566& 0.01565754&0.03246480&0.03296440&0.02521294&0.02694260\\
$\tau_{16}$ & 0.03031515& 0.01565469&0.03246373&0.03296242&0.02521179&0.02694084\\
$\tau_{15}$ & 0.02494569 &0.01423308&0.02715933&0.02765770&0.01905492&0.02088314\\
$\tau_{14}$ & 0.02447879 &0.01412248&0.02662948&0.02705660&0.01851858&0.02030185\\
$\tau_{13}$ & 0.02427643 &0.01407361&0.02641769&0.02682992&0.01830026&0.02007824\\
$\tau_{12}$ & 0.02409131 &0.01402853&0.02623127&0.02663907&0.01810142&0.01987946\\
$\tau_{11}$ & 0.02390885 &0.01398410&0.02604835&0.02645309&0.01790549&0.01968408\\
\midrule
\multicolumn{7}{c}{Put Option}\\
\midrule
$\tau_{20}$ & 0.04534067& 0.04044891&0.04765628&0.04838597&0.04477592&0.04625085\\
$\tau_{19}$ & 0.04136813& 0.01957849&0.04346118&0.04402728&0.03798709&0.03955459\\
$\tau_{18}$ & 0.04117866& 0.01857462&0.04320875&0.04371211&0.03762279&0.03914175\\
$\tau_{17}$ & 0.04116872& 0.01851743&0.04319187&0.04368645&0.03760252&0.03911518\\
$\tau_{16}$ & 0.04116820& 0.01851414&0.04319074&0.04368434&0.03760139&0.03911347\\
$\tau_{15}$ & 0.03572256& 0.01708147&0.03780731&0.03830289&0.03139350&0.03301665\\
$\tau_{14}$ & 0.03519281 & 0.01694566&0.03720163&0.03761288&0.3080348&0.03238388\\
$\tau_{13}$ & 0.03493847&0.01688151&0.03693765 &0.03733291&0.03052750&0.03210465\\
$\tau_{12}$ & 0.03470234& 0.01682184&0.03670090&0.03709169&0.03027123&0.03185029\\
$\tau_{11}$ & 0.03446938& 0.01676298&0.03646824&0.03685602&0.03001840&0.03159982\\
\bottomrule
\end{tabularx}
\end{table}

     \begin{table}[!ht]
\caption{Values of the callable and putable bond for initial short rate $r$ obtained by the eigenfunction expansion method}
\label{summarycallput}
\footnotesize
\begin{tabularx}{\textwidth} {XXXXXXX}
\toprule
 $r$& CIR& Vasicek& SubCIR, JD & SubCIR, PJ& SubVasicek, JD& SubVasicek, PJ\\
\midrule
0.01 &1.030391 &0.995407&1.054194&1.058549&1.022068&1.030678\\
0.02 &1.004673 &0.975223&1.025454&1.029652&0.998893&1.006540\\
0.03& 0.979637&0.955474&0.997443 &1.001420& 0.976211&0.982876\\
0.04& 0.955265&0.936150&0.970147&0.973843& 0.954015&0.959680\\
0.05& 0.931540&0.917242&0.943553&0.946911 &0.932295&0.936946\\
0.06& 0.908443&0.898741&0.917644&0.920614& 0.911044&0.914668\\
0.07& 0.885958&0.880639&0.892409&0.894942 & 0.890253&0.892840\\
0.08& 0.864068&0.862926&0.867831&0.869886&0.869914&0.871456\\
0.09& 0.842758&0.845594&0.843898&0.845435&0.850019& 0.850510\\
0.10& 0.822011&0.828635&0.820595&0.821579&0.830559&0.829996\\
\bottomrule
\end{tabularx}
 \end{table}

\end{document}